\documentclass[10pt,twocolumn,amssymb,superscriptaddress,aps,prd,nofootinbib,showkeys,floatfix,longbibliography]{revtex4-1}
\usepackage{amsmath}
\usepackage{amssymb}
\usepackage{graphicx}
\usepackage{hyperref}
\usepackage[usenames,dvipsnames]{xcolor}
\usepackage{slashed}
\usepackage{ulem}

\newlength{\colw}
\newcommand{\sss}{\scriptscriptstyle}

\newcommand{\half}{\frac{1}{2}}

\renewcommand{\Re}{\operatorname{Re}}
\newcommand{\order}{{\cal O}}
\newcommand{\One}{1\kern-4.5pt1}
\providecommand{\xmax}{\ensuremath{x_{\max}}}

\newcommand{\ncfg}{N_{\text{cfg}}}

\newcommand{\Dslash}{\slashed{D}}
\newcommand{\Dslashbak}{\overleftarrow{\Dslash}}
\newcommand{\pslash}{\slashed{p}}

\newcommand{\zmp}{Z^{\sss(+)}_m}

\definecolor{DarkGreen}{rgb}{0,0.5,0}

\newcommand{\cprev}{\cite{Skullerud:2000un,Skullerud:2001aw}}

\newcommand{\cRQCD}{\cite{Bali:2014gha}}

\renewcommand{\Re}{\mathfrak{Re}}
\newcommand{\Tr}{\operatorname{Tr}}
\newcommand{\Fig}[1]{\mbox{Fig.\,\ref{#1}}}
\newcommand{\Eq}[1]{\mbox{Eq.\,\eqref{#1}}}

\begin{document}

\title{Quark propagator with two flavors of \texorpdfstring{$\order(a)$}{O(a)}-improved Wilson fermions}

\author{Orlando Oliveira}
\affiliation{CFisUC, Dep.\ de F\'\i sica, Universidade de Coimbra, 
3004--516 Coimbra, Portugal.} 
\affiliation{Dep.\ de F\'\i sica, Instituto Tecnol\'ogico da Aeron\'autica, Centro T\'ecnico Aeroespacial, 12.228-900 S\~ao Jos\'e dos Campos, S\~ao Paulo, Brazil.}

\author{Paulo J.\ Silva}
\affiliation{CFisUC, Dep.\ de F\'\i sica, Universidade de Coimbra, 
3004--516 Coimbra, Portugal.} 

\author{Jon-Ivar Skullerud}
\affiliation{
Department of Theoretical Physics, National University of Ireland Maynooth,
Maynooth, County Kildare, Ireland.}
\affiliation{
School of Mathematics, Trinity College, Dublin 2, Ireland.}

\author{Andr\'e Sternbeck}

\affiliation{
Theoretisch-Physikalisches Institut, Friedrich-Schiller-Universit\"at
Jena, 07743 Jena, Germany.}

\date{Mai 9, 2019}

\begin{abstract}
We compute the Landau gauge quark propagator from lattice QCD with two
flavors of dynamical $\order(a)$-improved Wilson fermions.  The
calculation is carried out with lattice spacings ranging from 0.06\,fm
to 0.08\,fm, with quark masses corresponding to pion masses $m_\pi\approx420, 290$
and 150 MeV, and for volumes of up to $(4.5\,\mathrm{fm})^4$.  Our
ensembles allow us to evaluate lattice spacing, volume and quark mass effects.  
We find that the quark wave function which is suppressed in the infrared, is 
further suppressed as the quark mass is reduced, but the suppression is
weakened as the volume is increased. The quark mass function $M(p^2)$
shows only a weak volume dependence. Hypercubic artefacts beyond $O(a)$
are reduced by applying both cylinder cuts and H4 extrapolations. 
The H4 extrapolation shifts the quark wave function systematically upwards but
does not perform well for the mass function.
\end{abstract}

\pacs{11.15.Ha,12.38.Aw,21.65.Qr}
\keywords{quark propagator, Landau gauge, QCD}
      
\maketitle

\section{Introduction}
\label{sec:intro}

The quark propagator is one of the fundamental objects of QCD, and
contains information regarding several of the core nonperturbative
features of the theory, namely dynamical chiral symmetry breaking and
the absence of quarks from the physical spectrum (confinement).
Specifically, a non-vanishing mass function even in the limit of
vanishing bare quark mass is a direct sign of dynamical chiral
symmetry breaking, while the absence of asymptotic quark states can be
translated to an absence of real poles in the propagator, or
equivalently, the lack of a positive spectral representation (see, e.g., the discussions in \cite{Roberts:1994dr,Fischer:2006ub,Cloet:2013jya,Eichmann:2016yit}).

Lattice calculations provide us with an opportunity to study these
essentially nonperturbative aspects of the quark propagator.
Furthermore, first-principles lattice calculations may be used to
validate the assumptions used in other nonperturbative approaches such
as Dyson--Schwinger equations (DSEs) and functional renormalization
group (FRG) calculations, like the recent studies in  
\cite{Williams:2014iea,Williams:2015cvx,Cyrol:2017ewj,Aguilar:2018epe}).  

The quark propagator is a gauge dependent quantity, and hence requires
a choice of gauge condition.  The most commonly used gauge, both in
lattice and DSE or FRG calculations is the Landau gauge, but other
gauge conditions, including Coulomb gauge, maximal abelian gauge and
general covariant gauges have also been employed.

In the past, after some early studies using Wilson fermions
\cite{Becirevic:1999rv,Becirevic:1999kb,Skullerud:2000un,Skullerud:2001aw,Boucaud:2003dx},
most studies of the lattice Landau gauge quark propagator have used
staggered
\cite{Bowman:2002bm,Parappilly:2005ei,Bowman:2005vx,Furui:2005mp} or
overlap
\cite{Bonnet:2002ih,Zhang:2003faa,Zhang:2004gv,Kamleh:2004aw,Kamleh:2007ud}
fermions.  There have also been calculations using chirally improved
fermions \cite{Schrock:2011hq}, as well as twisted mass Wilson
fermions \cite{Blossier:2010vt,Burger:2012ti}. In \cite{August:2013jia} there 
is even a lattice study for adjoint fermions but for the gauge group SU(2).

Although calculations with the same quark content, but different
fermion discretizations, should agree in the continuum limit, at
finite lattice spacing the lattice artifacts may differ widely, and a
comparison between different actions remains an important tool on the
way to a controlled continuum extrapolation.
In this paper, we present a calculation of the quark propagator
using gauge configurations with $N_f=2$ flavors of $\order(a)$-improved 
Wilson fermion for nearly physical quark masses. In our calculation
the quark propagator
is also $\order(a)$-improved. Preliminary results were reported in~\cite{Oliveira:2016muq}.

The paper is organized as follows. In Section~\ref{sec:simulation}
we give the details of our lattice simulations, including the lattice
parameters, gauge fixing and extraction of form factors, and outline our
tree-level correction procedure.
Section~\ref{sec:results} discusses our lattice results.  In
Section~\ref{sec:artefacts} we discuss in some detail the hypercubic
artifacts beyond tree level, and how they may be brought under
control.  We end with a brief summary in Section~\ref{sec:summary}. 

\section{Simulation details}
\label{sec:simulation}

\subsection{Gauge ensembles}

For the computation of the quark propagator we take a subset of the
gauge ensembles generated by the Regensburg QCD (RQCD) collaboration (see, e.g., \cite{Bali:2012qs,Bali:2014gha,Bali:2014nma})
using $N_f=2$ nonperturbatively improved Sheikholeslami--Wohlert (clover)
fermions \cite{Sheikholeslami:1985ij} and the Wilson action for the gauge sector. In the present work we use three values 
for the gauge coupling, corresponding to lattice spacings of $a\approx0.081\,\text{fm}$, 
$a\approx0.071\,\text{fm}$ and $a\approx0.060\,\text{fm}$, and quark masses corresponding to 
pion masses of $m_\pi\approx420\,\text{MeV}$, $m_\pi\approx290\,\text{MeV}$, and 
$m_\pi\approx150\,\text{MeV}$, which is almost at the physical point 
\cite{Bali:2012qs,Bali:2014nma}. Most of the calculations
have been carried out on a lattice volume of $32^3\times64$, but 
the near-physical quark mass ensemble was generated on a larger $64^4$
lattice, corresponding to a physical volume of $(4.5\,\textrm{fm})^4$.  We
have also used a $64^4$ lattice to check finite volume effects for one
of the other parameter choices ($m_\pi\approx290$ MeV).
The parameters used are listed in Table~\ref{tab:params}.  

\begin{table}
\begin{tabular}{r@{\quad}c@{\quad}c@{\quad}c@{\quad}c@{\quad}c@{\quad}c}
\hline\hline
 no. &$\beta$ & $\kappa$  & $V$ & $a$ [fm] & $(m_\pi,m)$ [MeV] &  $\ncfg$ \\ \hline
I & 5.20 & 0.13596 & $32^3\!\times\!64$ & 0.081 & $(280,~6.2)$ 
  & 900 \\*[0.5ex]  
II & 5.29 & 0.13620 & $32^3\!\times\!64$ & 0.071 & $(422,17.0)$ 
& 900 \\  
III & 5.29 & 0.13632  & $32^3\!\times\!64$ & 0.071& $(295,~8.0)$      & 908 \\      
IV  & 5.29 & 0.13632  & $64^3\!\times\!64$ & 0.071& $(290,~8.0)$      & 750 \\      
V &  5.29 & 0.13640   & $64^3\!\times\!64$ &  0.071& $(150,~2.1)$  & 400 \\*[0.5ex] 
VI & 5.40 & 0.13647  & $32^3\!\times\!64$ & 0.060 & $(426,18.4)$ 
 & 900 \\
\hline \hline
\end{tabular}
\caption{Lattice parameters used in this study.  The lattice spacings
  $a$ and pion masses $m_\pi$ are taken from \cRQCD, while
  $am=1/(2\kappa)-1/(2\kappa_c)$ is the subtracted bare quark mass
  obtained using the critical hopping parameters $\kappa_c$ from \cRQCD.}
\label{tab:params}
\end{table}

For the gauge fixing we used an over-relaxation algorithm which iteratively 
maximizes the Landau-gauge functional
\begin{equation}
 F_U[g] = \frac{1}{4V}\sum_{x\mu} \Re\Tr U^{g}_{x\mu}
\end{equation}
with $U^{g}_{x\mu} = g_x U_{x\mu} g^{\dagger}_{x+\hat{\mu}}$ and $g_x\in SU(3)$.
As stopping criterion we used
\begin{equation}
 \max_{x}\; \Re\Tr\left[(\nabla_\mu A_{x\mu})(\nabla_\mu 
  A_{x\mu})^{\dagger}\right] < 10^{-9}
\end{equation}
where $A_{x\mu} \equiv \frac{1}{2iag_0}(U^g_{x\mu}-U_{x\mu}^{g\dagger}) \vert_{\mathrm{traceless}}$ and 
$\nabla_\mu A_{x\mu} \equiv \sum_\mu (A_{x\mu} - A_{x-\hat{\mu},\mu})$, as usual. 
This stopping criterion corresponds to demanding an average value over the lattice of
$\Re\Tr\left[(\nabla_\mu A_{x\mu})(\nabla_\mu A_{x\mu})^{\dagger}\right]  \lesssim 10^{-12}$. 
In this study, we do not attempt to investigate the influence of Gribov copies on the quark propagator.

\subsection{\texorpdfstring{$\boldsymbol{O(a)}$}{O(a)}-improved Wilson quark propagator}

For the Wilson-clover action, the $\order(a)$-improved quark propagator is given
by \cite{Sheikholeslami:1985ij,Heatlie:1990kg,Skullerud:2000un,Skullerud:2001aw}
\begin{equation} 
 S_{\mathrm{rot}}(x,y) =  (1+2b_q am) \left\langle L(x) M^{-1}_{SW}(x,y) R(y) \right\rangle_U \; ,
\label{eq:rotprop}
\end{equation}
where $M^{-1}_{SW}$ is the inverse Wilson clover-fermion matrix
which is rotated from left and right by\footnote{For the covariant derivative we use
\begin{subequations}
\begin{align} 
 \overrightarrow{\Dslash}(x) \psi (x) &\equiv \sum_\mu\frac{\gamma_{\mu}}{2a}\,  \left[ U_{x\mu} \psi(x+\mu) - U^{\dagger}_{x-\hat{\mu},\mu} \psi(x-\mu) \right]\,, \\
  \bar{\psi}(x) \Dslashbak(x) &\equiv \sum_\mu \left[ \bar{\psi}(x+\hat{\mu}) U^{\dagger}_{x\mu} -  \bar{\psi}(x-\hat{\mu}) U_{x-\hat{\mu},\mu}  \right] \, \frac{\gamma_{\mu}}{2a} \,.
\end{align}
\end{subequations}
}
\begin{subequations}
\begin{align}
 L(x) &\equiv \left[1-c_q a\overrightarrow{\Dslash}(x)\right]\;, \\
 R(y) &\equiv \left[1+c_q a\Dslashbak(y)\right]\,.
\end{align}
\end{subequations}
The improvement coefficients $b_q$ and $c_q$ should be nonperturbatively 
determined to remove the $\order(a)$ errors in the quark propagator completely. We expect 
however that the deviations from tree level are small and therefore fix them at 
their tree-level (tl) values, i.e., $b_q=c_q=1/4$ \cite{Heatlie:1990kg}. 
On the lattice, the bare quark mass is
\begin{equation}
 am = \frac{1}{2}\left(\frac{1}{\kappa}-\frac{1}{\kappa_c(\beta)}\right)\quad
 \xrightarrow{\beta\to\infty}\; \frac{1}{2\kappa}-4 \equiv am_0\, ,
\end{equation}
where $\kappa_c(\beta)$ is the critical hopping parameter. This tl-rotated quark propagator 
is the one used throughout in this paper.
For its computation we use correspondingly rotated point sources at four different 
source locations (except for the $64^4$ lattices, where only two sources were 
used) and average the data from the different sources.

Note that in \cprev, a different improved propagator was also analyzed, but as 
it was found that this propagator has more severe lattice artifacts beyond 
$\order(a)$, we will not consider it here.

\subsection{Lattice tree-level corrections}
\label{sec:corrections}

In the continuum, the renormalized Euclidean-space vacuum quark propagator can be
written as $S^{ab}(p,\mu)=\delta^{ab}S(p,\mu)$ with
\begin{equation}
S(p,\mu) = \frac{1}{i\pslash A(p^2,\mu) + B(p^2,\mu)} \equiv
\frac{Z(p^2,\mu)}{i\pslash +M(p^2)}\,,
\end{equation}
where $\mu$ is the renormalization scale.
The propagator is completely determined by the two form factors $A(p^2,\mu)$ 
and $B(p^2,\mu)$ or alternatively $Z(p^2,\mu)$ and $M(p^2)$, the quark wave 
and mass function, respectively. Due to multiplicative renormalizability
$M$ is renormalization-group invariant (see, e.g., \cite{Skullerud:2000un,Pennington:2005be}) and in momentum subtraction (MOM) schemes is set equal to the running quark mass $M(\mu^2)=m_R(\mu)$ at some ultraviolet renormalization scale, while $Z(\mu^2,\mu)=1$.

On a finite lattice with periodic boundary conditions in space and
antiperiodic boundary conditions in time for the fermions, the available 
momenta are discrete and given by 
\begin{subequations}
\begin{align}
p_i & = \frac{2\pi}{N_i a}\left(n_i - \frac{N_i}{2}\right)\;;
 & 
 n_i = & 1,2,\cdots,N_i\,, \\
p_t & = \frac{2\pi}{N_t a} \left(n_t-\half-\frac{N_t}{2}\right)\;;
& ,
 n_t = & 1,2,\cdots,N_t \,,
\label{eq:latt_momenta}
\end{align}
\end{subequations}
where $N_i$, $N_t$ are the number of lattice points in the spatial and
temporal directions, respectively. The lattice quark propagator can still be 
parametrized by two form factors, for instance $A_{L}(p,a)$ and $B_{L}(p,a)$
which in the asymptotic limit and after renormalization reduce to the corresponding 
continuum functions, i.e., 
\begin{subequations}
 \begin{align}
      A(p^2,\mu) &= \lim_{a\to0} Z_2(\mu,a) A_{L}(p,a) \,, \\ 
      B(p^2,\mu) &= \lim_{a\to0} Z_2(\mu,a) B_{L}(p,a) \,,
 \end{align}
\end{subequations}
with the quark wave renormalization constant $Z_2(\mu,a)$. Note that
for any finite lattice spacing, $A_L(p,a)$ and $B_L(p,a)$ are not
functions of $p^2$ alone, but also depend on all the other invariants
of the (hypercubic) $H(4)$ symmetry group. We therefore keep $p$
instead of $p^2$ in the argument of the lattice form factors. For
larger $p$, especially if $ap$ is large, the lattice data will expose
a different momentum dependence, aka.\ ``fish-bone structure'' which
survives the renormalization. Only in the asymptotic limit, where the
lattice spacing is sufficiently small such that the lattice structure
is irrelevant, do these deviations disappear for any
fixed value of the momentum $p$. Improved lattice operators and actions help to improve the convergence towards the continuum.

For commonly used values of $\beta$ these deviations due to the lattice spacing
are large when using an unimproved (clover-) Wilson fermion propagator. We therefore use the 
$\order(a)$-improved (tl-rotated) lattice Wilson fermion propagator given in
Eq.\,\eqref{eq:rotprop}, which at tree level reads \cite{Skullerud:2000un}
\begin{subequations}
\begin{align}
 S^{(0)}_{\mathrm{rot}}(p,a,m) &= \frac{Z_{\mathrm{rot}}^{(0)}(p,a,m)}{i\slashed{\bar{p}}+ M_{\mathrm{rot}}^{(0)}(p,a,m)} \, ,
  \label{eq:Srot0_ZM}
\intertext{with}
  Z_{\mathrm{rot}}^{(0)}(p,a,m) &= \left(1 + \frac{am}{2}\right)\frac{\bar{p}^2 I_A^2 + I_B^2}{I_A D_W} \,, \\
  M_{\mathrm{rot}}^{(0)}(p,a,m) &= \frac{I_B}{I_A}
\end{align}
\end{subequations}
and
\begin{subequations}
\begin{align}
 D_W &\equiv \bar{p}^2 + [m + M_W]^2 \,,\\
 I_A &\equiv 1 + \frac{am}{2} + \frac{3a^2\bar{p}^2}{16} + \frac{a^4 \left(\hat{p}^2-\bar{p}^2\right)}{4} \,, \\
 I_B &\equiv m\left(1 - \frac{a^2\bar{p}^2}{16}\right) - \frac{a^3}{32}\bar{p}^2\hat{p}^2 + \frac{a^3 \left(\hat{p}^2-\bar{p}^2\right)}{2}\,.
\end{align}
\end{subequations}
The Wilson mass term $M_W=a\hat{p}^2/2$ and the lattice momentum functions are
\begin{align}
 \label{eq:pbar}
 \bar{p}_\mu &\equiv \frac{1}{a}\sin(ap_\mu) \,,  \\
 \label{eq:phat}
 \hat{p}_\mu &\equiv \frac{2}{a}\sin(ap_\mu/2)\,.
\end{align}
Knowledge of the lattice tree-level propagator is of great advantage when one extracts the nonperturbative quark form factors from the lattice data. With appropriate projections one obtains form factors which are exact at tree level. We will see this reduces a large fraction of the lattice artifacts mentioned above. 

Following this, we define the (renormalized) nonperturbative
quark wave and mass functions as
\begin{align} 
\label{eq:renZ}
 Z(p,\mu,m) &= Z_2^{-1}(\mu,a) \cdot Z_L(p,a,m) \\ 
 Z_L(p,a,m) &= Z_{\mathrm{rot}}(p,a,m) / Z_{\mathrm{rot}}^{(0)}(p,a,m) \intertext{and}
 M_L(p,a,m) &= \Big(M_{\mathrm{rot}}(p)-M_{\mathrm{rot}}^{(-)}(p)\Big)\,/\,\zmp(p)
\label{eq:hybridM}
\end{align}
where $M_{\mathrm{rot}}^{(0)}$ has been split into strictly positive and negative terms \cite{Skullerud:2001aw}:
\begin{equation}
 \label{eq:hybridM_coeff}
  M^{(0)}_{\mathrm{rot}}(p) \equiv m \underbrace{Z^{(+)}_m(p)}_{\ge 1} + \underbrace{M_{\mathrm{rot}}^{(-)}(p)}_{\le 0}\,.
\end{equation}
$Z_L$ and $M_L$ have the correct form at tree level if $Z_{\mathrm{rot}}= 1/A_{\mathrm{rot}}$ and $M_{\mathrm{rot}}=B_{\mathrm{rot}}/A_{\mathrm{rot}}$ are obtained from the usual traces of the lattice data for $S_{\mathrm{rot}}$ [\Eq{eq:rotprop}]:
\begin{subequations}
\label{eq:ALBL}
\begin{align}
 A_{\mathrm{rot}}(p,a,m) &= \frac{1}{12\bar{p}^2}\Tr  \left[\slashed{\bar{p}} ~S^{-1}_{\mathrm{rot}}(p,a,m)\right]\;, \\
 B_{\mathrm{rot}}(p,a,m) &= \frac{1}{12}\Tr \left[S^{-1}_{\mathrm{rot}}(p,a,m)\right] \,.
\end{align}
\end{subequations}
with $\bar{p}$ defined in \Eq{eq:pbar}. In the continuum, the mass function is renormalization-group invariant. 
Hence $M(p^2,m)=\lim_{a\to 0} M_L(p,a,m)$. We find only a very weak dependence on the lattice spacing in our data, and hence conclude that for our lattice spacings we are close to the limit. To simplify
notation, we therefore drop the subscript $L$, when showing lattice data for the mass function.

\begin{figure}[tb]
 \includegraphics[width=\linewidth]{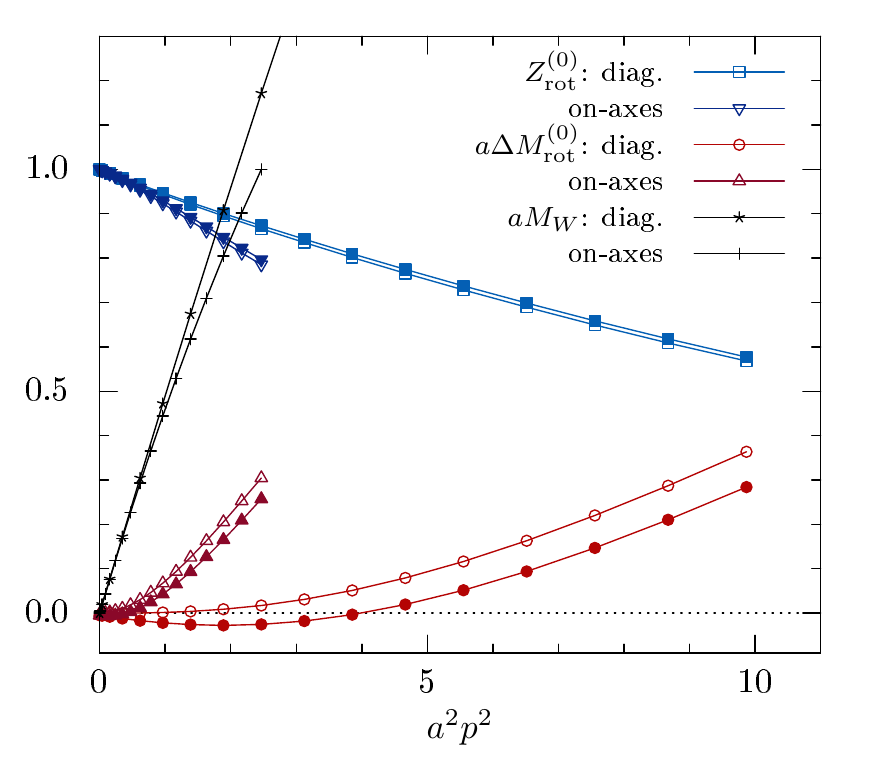}
 \caption{$Z_{\mathrm{rot}}^{(0)}(p)$, $aM_{W}(p)$ and 
  $a\Delta M_{\mathrm{rot}}^{(0)}(p)\equiv aM_{\mathrm{rot}}^{(0)}(p)-am$ versus $a^2p^2$ for diagonal and on-axes momenta $ap$.
   Open symbols are for $am=0$, full symbols for $am=0.1$.}
 \label{fig:Z0M0}
 \includegraphics[width=\linewidth]{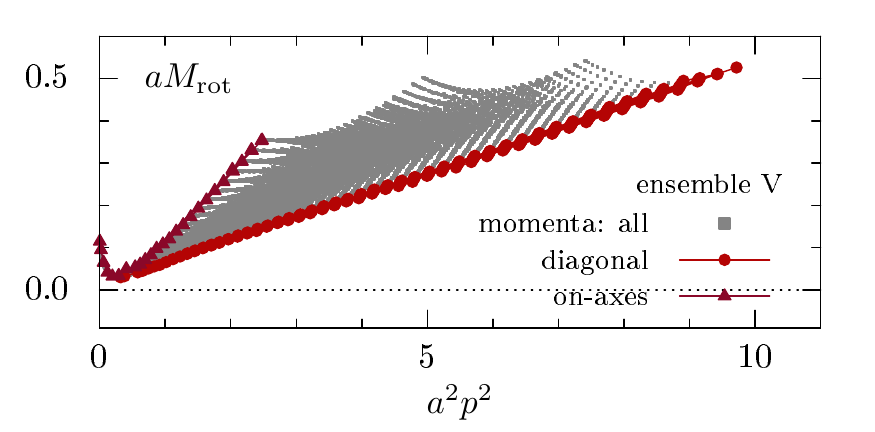}
 \caption{Bare uncorrected lattice data for $aM_{\mathrm{rot}}(p)$ versus $a^2p^2$ for ensemble V (see Table \ref{tab:params}). Circles are for (near) diagonal momenta and triangles for (near) on-axis momenta.}
 \label{fig:M_uncorr_b5p29kp13640_64x64}
\end{figure}

Here a note on our lattice tree-level corrections of the mass function is in order: Naively one would correct it by subtracting the $p$-dependent part of the lattice tree-level expression, i.e.,
\begin{equation}
  M_{\mathrm{sub}}(p) = M_{\mathrm{rot}}(p) - M_{\mathrm{rot}}^{(0)}(p,m) + m\,,
   \label{eq:M_add_treel}
\end{equation}
analogously as one would subtract $M_W$ in case of the unimproved Wilson quark propagator. Alternatively, one could apply a multiplicative correction, i.e., define the nonperturbative mass function as
\begin{equation}
  M_{\mathrm{mul}}(p) = M_{\mathrm{rot}}(p)/Z^{(0)}_m(p)
\end{equation}
with $M_{\mathrm{rot}}^{(0)}(p)\equiv m\cdot Z^{(0)}_m(p)$. However, this
approach suffers from cancellation effects when either $M_{\mathrm{rot}}$ or $M_{\mathrm{rot}}^{(0)}$ take values around zero (cf.~\Fig{fig:Z0M0}). It was found in \cite{Skullerud:2001aw} 
that the ``hybrid'' tree-level correction in Eqs.\,\eqref{eq:hybridM} and \eqref{eq:hybridM_coeff} combines the advantages of the additive and multiplicative correction. Hence we use it for the results for $M$ shown below. A comparison with $M_{\mathrm{mul}}(p)$ is shown in the appendix.

For illustrative purposes, we show in \Fig{fig:Z0M0} the momentum dependence of 
$Z_{\mathrm{rot}}^{(0)}$, $a\Delta M_{\mathrm{rot}}^{(0)} \equiv aM_{\mathrm{rot}}^{(0)}-am$
and the Wilson mass 
term $aM_W$, both for diagonal and on-axes momenta $ap$. Filled symbols are for
$am=0.1$, open symbols for $am=0$. We see that $a\Delta M_{\mathrm{rot}}^{(0)}$
changes sign depending on $ap$ and $am$, while $Z_\mathrm{rot}$ is almost unaffected by the quark mass.
Also note that the $\order(a)$ improvement is already evident at tree-level: 
$a\Delta M^{(0)}_{\mathrm{rot}}$ increases much less with $ap$ than $aM_W$ which 
dominates the lattice artifacts of the unimproved Wilson (clover-) fermion propagator. 
On-axes momenta cause 
larger effects, than momenta along diagonal lattice directions, for which
$aM_{\mathrm{rot}}^{(0)}\simeq am$ even up to $ap<2$. Changing the quark mass 
changes $a\Delta M_{\mathrm{rot}}^{(0)}$ much less than the momentum orientation. 
Note that for our $\beta$ values, $am=0.1$ corresponds to larger quark 
masses than we use for our study. Our values for $am$ range from $am = 0.00075$ (ensemble V) 
to $am=0.006135$ (ensemble II). The value of $am$ is hence negligible for the tree-level corrections; only the type momentum orientation matters. 

In \Fig{fig:M_uncorr_b5p29kp13640_64x64} we show, as an example, bare 
uncorrected lattice data for $aM_{\mathrm{rot}}$ for our lightest-quark ensemble V (aka.\ fishbone plot). 
The points scatter depending on the size and type of momentum; on-axes momenta show 
the largest and diagonal momenta the smallest discretization effects. Comparing \Fig{fig:Z0M0}
and \ref{fig:M_uncorr_b5p29kp13640_64x64} we see the dominant part of these 
effects is already contained in the tree-level propagator. This
explains the effectiveness of applying lattice tree-level corrections, though we
also see that the linear $a^2p^2$ dependence sets in at much lower $a^2p^2$ than it does 
for the tree-level curves.

\subsection{Beyond lattice tree-level corrections}

Using the tl-rotated quark propagator and the tree-level correction helps to
drastically reduce the discretization effects. The figures
below will clearly evidence this. However, the discretization artifacts are not 
removed completely. To reduce the remaining hypercubic artifacts, we consider two strategies:
\begin{enumerate}
\item \underline{Data cuts:}  We have employed the cylinder cut first described in
  \cite{Leinweber:1998uu} to select momenta close to the diagonal in 4-momentum
  space, which have the smallest hypercubic artifacts.  We have also
  considered an alternative cut based on the value of $x\equiv
  p^{[4]}/p^{[2]}$ (see below), selecting only momenta for which
  $x<\xmax$.  The results of this cut are similar to those of the
  cylinder cut, but the cylinder cut gives a more even distribution of
  points across the entire momentum range and is hence preferred.
\item \underline{H4 extrapolation~\cite{Becirevic:1999uc,deSoto:2007ht}:}  This
  will be described in more detail in Section~\ref{sec:artefacts} below.
\end{enumerate}

\section{Results}
\label{sec:results}

\subsection{Tree-level corrected data}

\begin{figure*}
\mbox{\includegraphics[height=7cm]{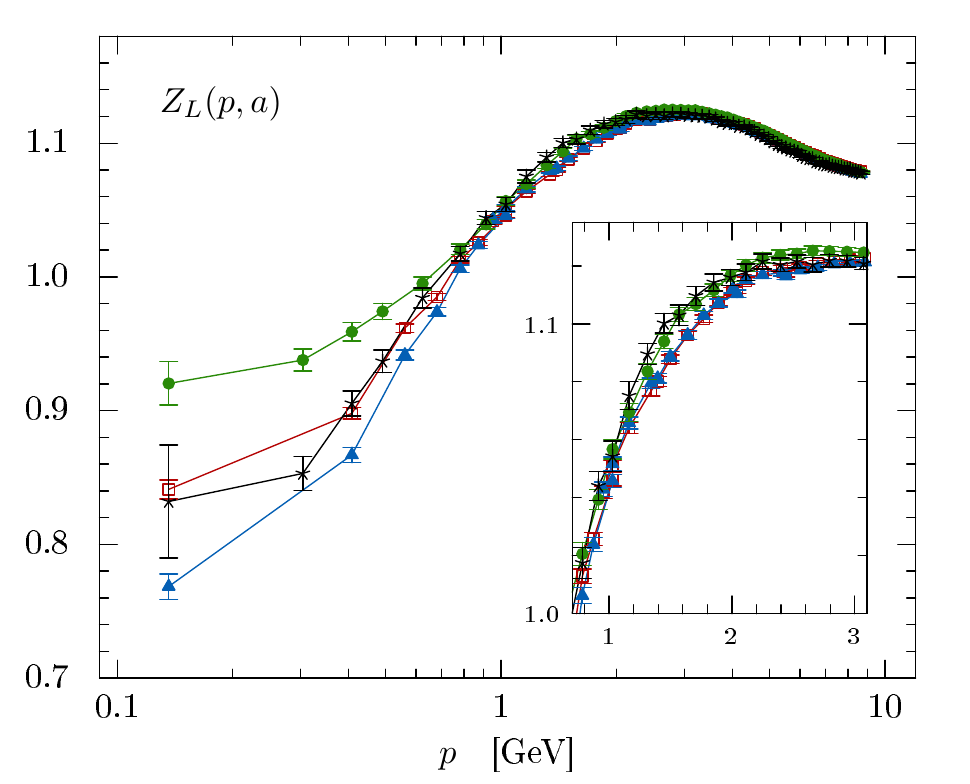}
\includegraphics[height=7cm]{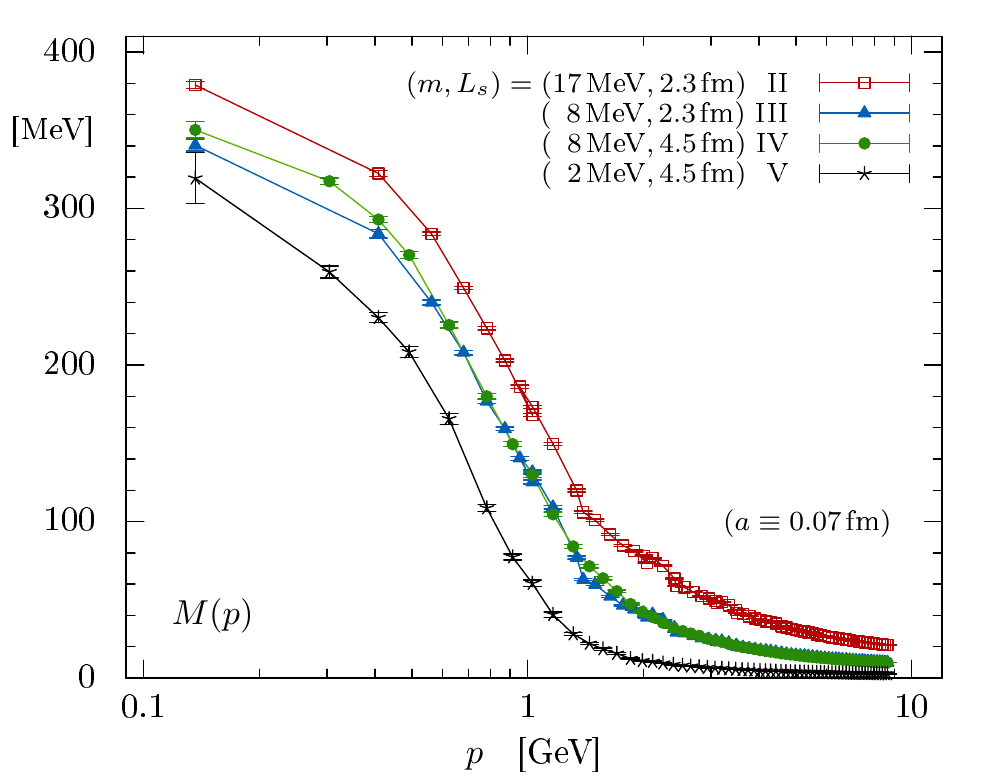}}
\caption{The unrenormalized tree-level corrected wave function $Z_L(p,a)$ 
  (left) and the hybrid tree-level corrected mass function $M_L(p)$ (right) as a 
  function of $p$. The data are for a fixed lattice spacing 
  ($\beta=5.29$), but different spatial extent $L_s$ and quark masses $m$
  to demonstrate volume and quark mass effects. The corresponding gauge field ensembles are 
  II, III, IV and V. Momenta are cylinder 
  cut with a radius of 1 lattice momentum unit. The top right legend applies to both plots.}
\label{fig:Z-volM-vol}
\bigskip
\mbox{\includegraphics[height=7cm]{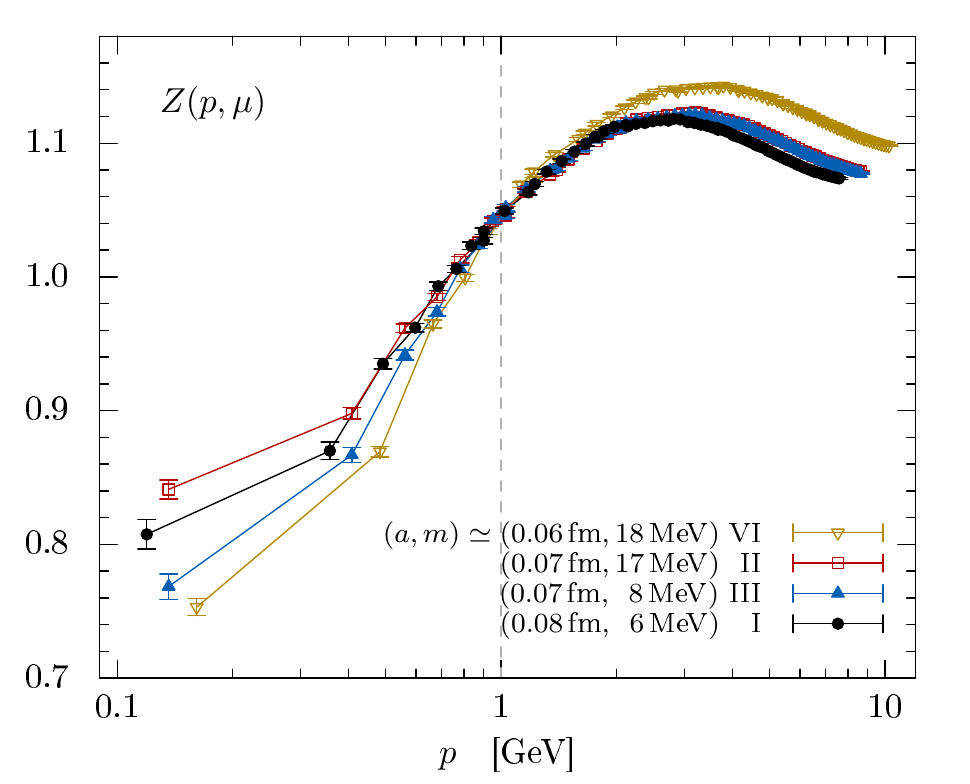}
\includegraphics[height=7cm]{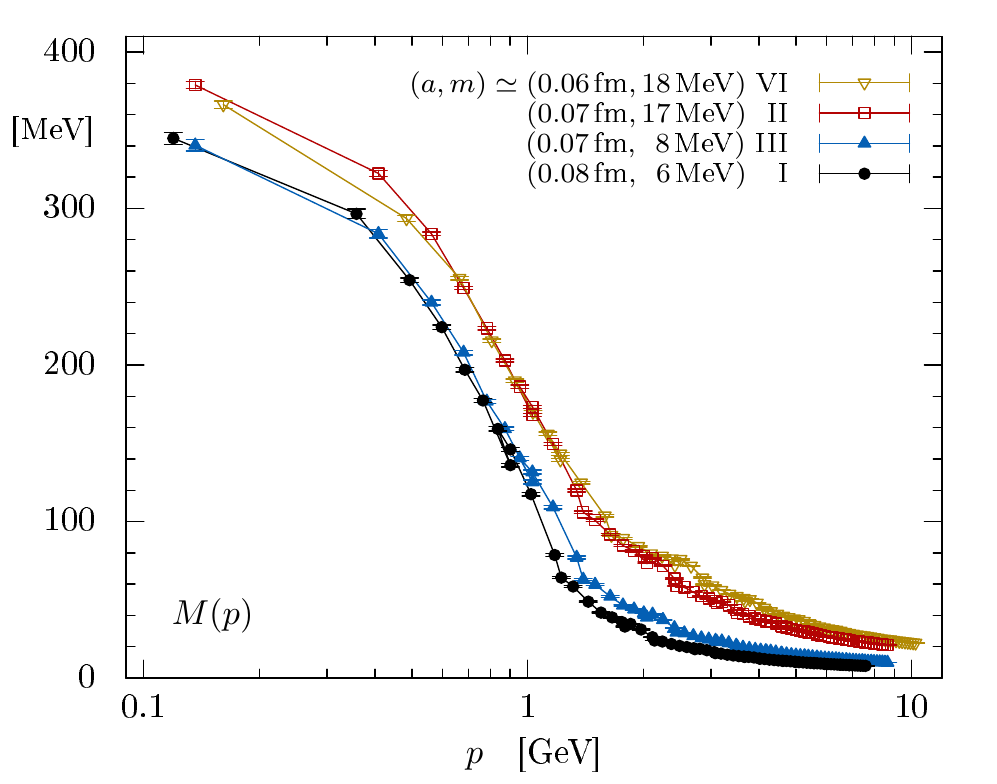}}
\caption{The renormalized tree-level corrected wave function $Z(p,\mu)$ 
  (left) and the hybrid tree-level corrected mass function $M(p)$ (right) as a 
  function of $p$, for our four ensembles on a 
  $32^3\times64$ lattice (I, II, III and VI). They have been renormalized 
  at $\mu=1\,\textrm{GeV}$, relative to the $\beta=5.29$ points 
  ($a=0.07\,\textrm{fm}$). Open symbols refer to data for 
  $m\simeq$ 17--18~MeV; full symbols for 
  $m\simeq$ 6--8~MeV.}
\label{fig:Z-allM-all}
\end{figure*}

We start our discussion of data with the four ensembles at $\beta=5.29$ and 
focus first on volume and quark mass effects. The tree-level corrected results
for $Z(p,a)$ and $M(p)$ are shown in Fig.\,\ref{fig:Z-volM-vol} as a function
of $p$. The quark wave function is left unrenormalized ($Z_2\equiv 1$), because 
the data points are for a single $\beta$. As expected, for large momenta the 
quark mass dependence of $Z$ is negligible; the points for $Z$ almost collapse 
onto a single curve. For $p<3\,\text{GeV}$, however, deviations grow as $p$ 
decreases. Between $p=1$ and 2 GeV, both a larger spatial volume and a smaller 
quark mass value cause points to move up. Interestingly, around 
$p=1\,\textrm{GeV}$ points for the different sets almost coincide, 
although no renormalization was applied. For $p<1\,\textrm{GeV}$ 
deviations grow again towards the infrared, depending on quark mass and volume: 
at fixed $p$, a larger volume causes $Z$ to move up [compare triangles and circles in \Fig{fig:Z-volM-vol}
(left)], while a smaller quark mass causes the opposite effect (compare 
circles to crosses, or squares to triangles). Within our parameter ranges, the 
quark mass suppression is similar in size to the enhancement with volume.

For the mass function $M(p)$ at $\beta=5.29$ (\Fig{fig:Z-volM-vol}, right)
we see a clear quark mass dependence. Varying $m$ not only changes the offset 
at large $p$, but also the functional form of $M(p^2)$. A simple rescaling of 
$M(p^2)$ or a subtraction of a finite offset will not collapse the data 
points onto a single curve.\footnote{A simple rescaling yields curves for 
$M(p)$ which are suppressed at low $p$, the more the larger $m$ is. On the other 
hand, when subtracting a finite offset, the curves coincide at 
$p\simeq 5\,\textrm{GeV}$ and approximately also at small $p$ but not in between.}
The volume effect for $M(p)$ is small in comparison and actually only resolvable for 
$p<0.6\,\textrm{GeV}$. A larger volume causes points to move slightly up 
(compare circles to triangles for $p<0.6\,\textrm{GeV}$).

Next we look at discretization effects for which we compare our $32^3\times 64$ 
data for $\beta=5.20$, 5.29 and 5.40 (ensembles I, II, III and VI). We have to 
apply a renormalization factor $Z_2$, separately for each $\beta$ (see \Eq{eq:renZ}). 
For a better comparison with \Fig{fig:Z-volM-vol} we again set $Z_2\equiv 1$ 
for $\beta=5.29$ and renormalize the other two sets (I and IV) 
relative to that. As renormalization point we chose $\mu=1\,\textrm{GeV}$ for
which we found the smallest volume and quark mass effects at small $p$. 
For the same reason we could chose any other point above 
$3\,\textrm{GeV}$ as well, but for large $p$ we actually expect (and find) 
discretization effects. Renormalizing there would artificially shift these 
effects to smaller momenta where they would overlap with volume and quark 
mass effects. Choosing $\mu=1\,\textrm{GeV}$ is thus optimal
for our purposes.

$M$ was not renormalized, because it is renormalization group invariant if 
lattice discretization effects are removed. We will now analyze these effects. 

Our results for $Z(p,\mu)$ and $M(p)$ are shown in \Fig{fig:Z-allM-all} and 
we clearly see discretization effects for larger~$p$. In particular,
the non-monotonic behavior of $Z(p,\mu)$, reaching a maximum at
$p\sim3$~GeV and bending down towards larger~$p$, is an effect seen in
previous studies with Wilson--clover fermions which is absent in
studies using other discretizations. By looking at the bare uncorrected data 
(not shown) we find that the tree-level correction, in combination with the momentum 
selection (cylinder cut), indeed removes most of the discretization effects.
This removal is not complete, as expected, and what remains is seen in 
Fig.\,\ref{fig:Z-allM-all}. If the removal was complete, the points for $Z$ 
above 1\,GeV would collapse onto a single curve and only at small 
momentum would deviations due to volume or quark mass effects be seen.

\begin{figure*}
\centering
\hbox{\includegraphics[height=6.2cm]{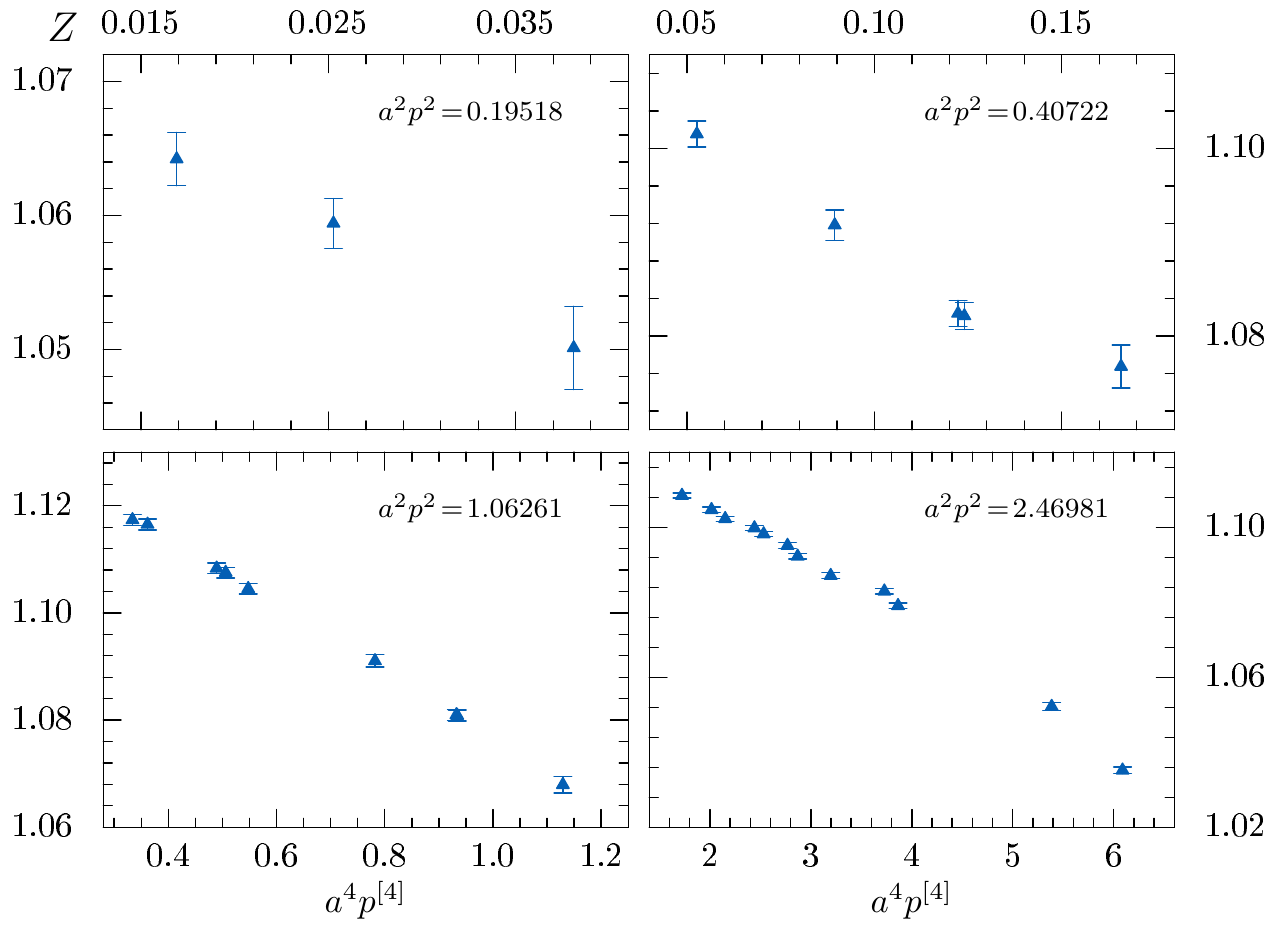} \quad
  \includegraphics[height=6.2cm]{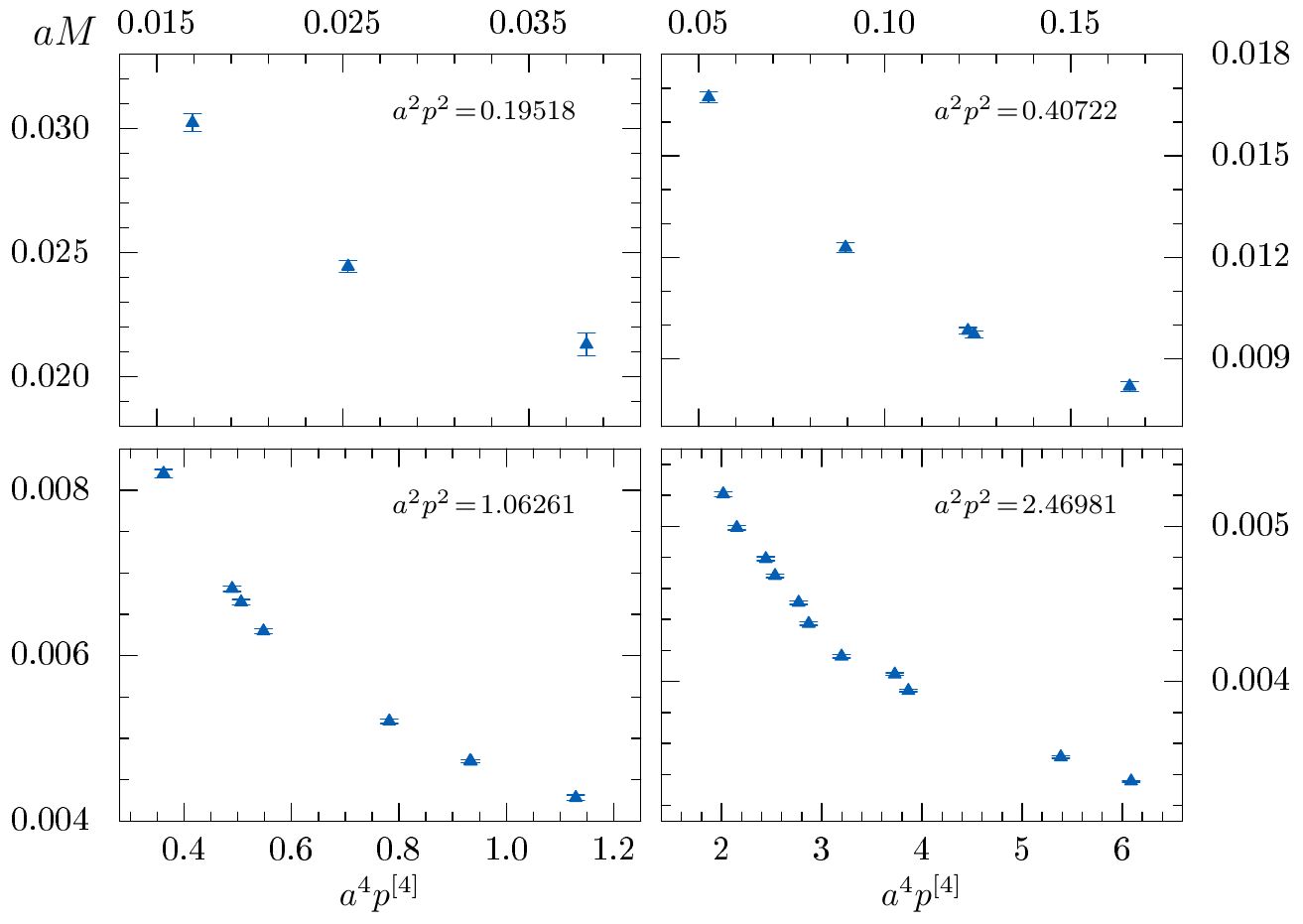}}
\hbox{\includegraphics[height=6.2cm]{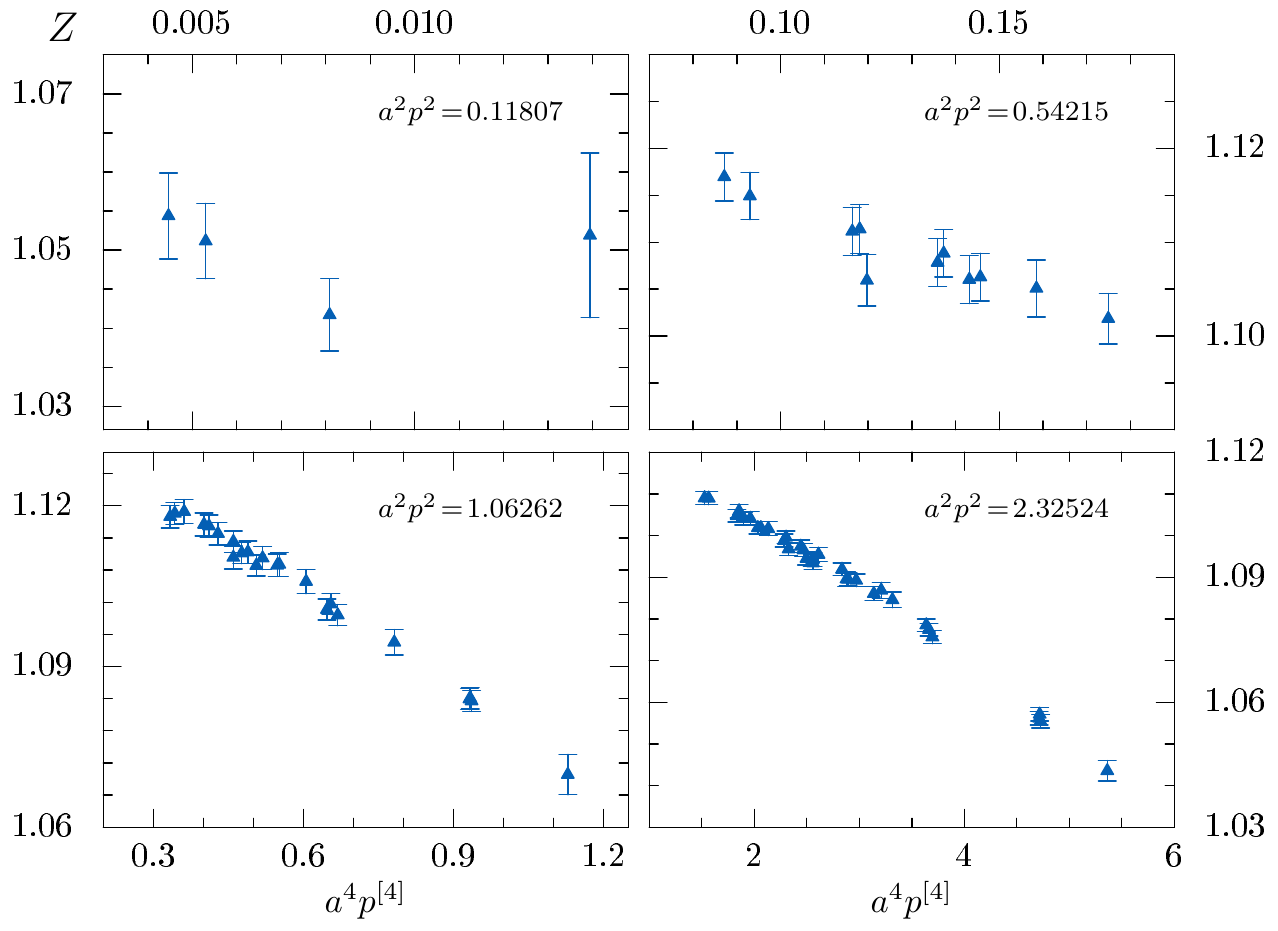} \quad
 \includegraphics[height=6.2cm]{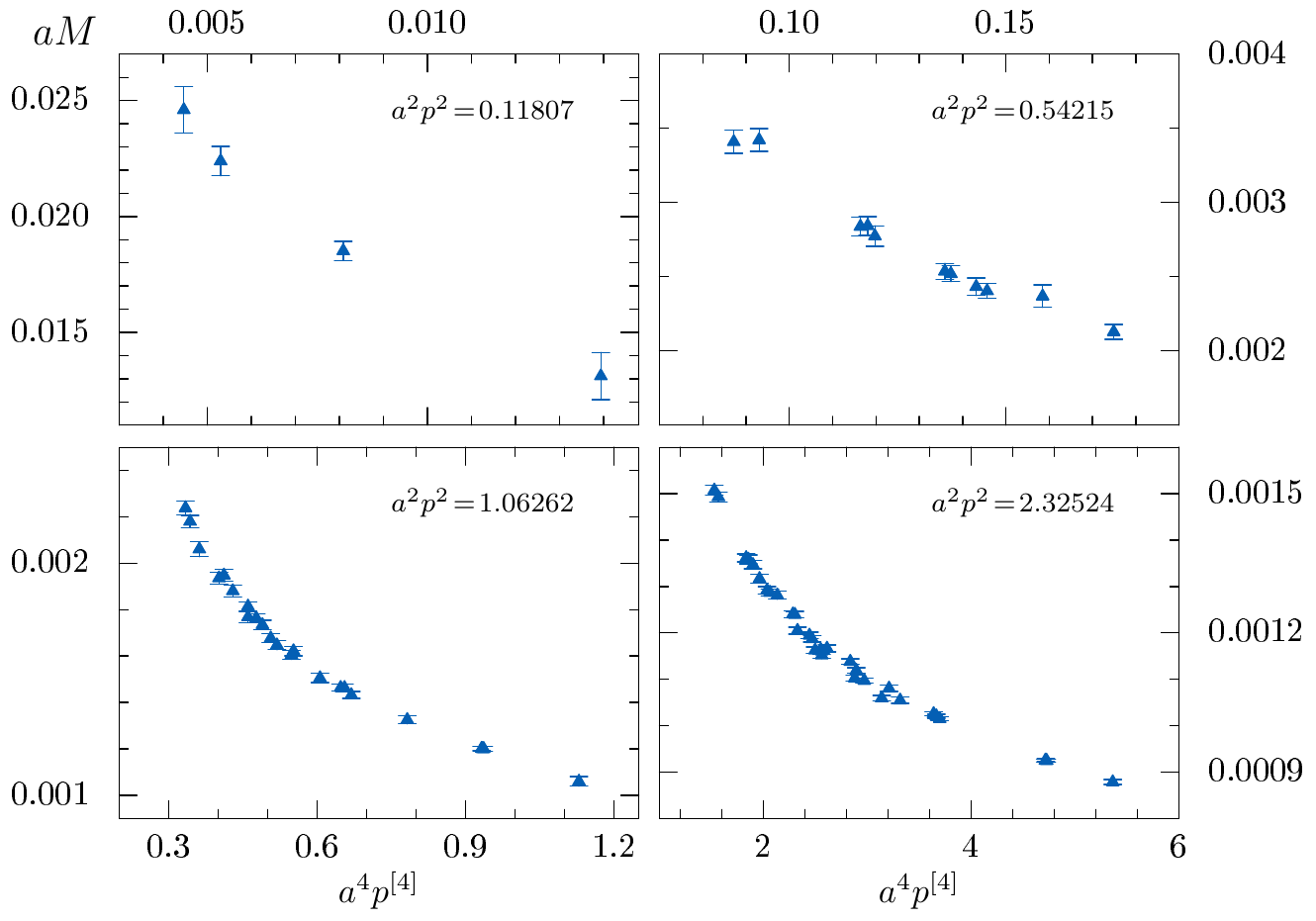}}
\caption{$Z_L(p^2,p^{[4]})$ (left panels) and $aM_L(p^2,p^{[4]})$ (right panels) as 
   a function of $a^4 p^{[4]}$ at four values of $a^2 p^2$. The top
   panels show data for $\beta=5.29$, $\kappa = 0.13632$ and a $32^3\times64$ lattice, 
   the bottom panels for $\beta=5.29$, $\kappa = 0.13640$ and a $64^4$ lattice.}
\label{fig:ZMp2H4_b5p29kp13632_kp13640}
\end{figure*}

Similarly, for $M(p)$ the renormalization group invariance is broken
by lattice artifacts. In Fig.\,\ref{fig:Z-allM-all} (right) 
we see that the data for $M$ with approximately the same $m$ overlap within errors 
for $p<1\,\textrm{GeV}$, while for $p>1$ GeV a similar but slightly different $p$-dependence
is seen. For the reader's convenience we have used open and 
full symbols in \Fig{fig:Z-allM-all} to indicate the respective~$m$. Overall,
lattice spacing effects for $M$ are smaller than for $Z$.

It is indeed reassuring to see that the bending down of $Z(p,\mu)$ sets in at higher $p$ the finer the lattice (compare the points for ensemble I and VI in \Fig{fig:Z-allM-all}). Also, the mass function $M(p)$
falls off for large $p$ such that one can assume that it will approach 
the perturbative running of the quark mass in the ultraviolet limit if all 
discretization effects are subtracted. In the infrared momentum limit we see the dynamically generated ``constituent'' quark mass of about 300--400 MeV, which one would expect. It has been
suggested \cite{Fischer:2003rp,Aguilar:2010cn} that $M(p)$ should reach a plateau 
at small $p$. With our data we can neither confirm nor refute this.  We see a slight change of slope at small $p$ for the ensembles III and~IV, but data points for much lower $p$ are needed to address this.

\subsection{Correction of hypercubic artifacts}
\label{sec:artefacts}

Using the tl-rotated quark propagator and the tree-level correction described 
above, we obtained quark dressing functions for cylinder-cut momenta which show 
much smaller lattice spacing effects than unimproved and uncorrected data for the clover
quark propagator would show. However, the discretization effects are not removed completely. 
We will now attempt to reduce the remaining hypercubic discretization effects 
using the (so-called) H4 method~\cite{Becirevic:1999uc,deSoto:2007ht}. Note 
that our implementation differs slightly from the original proposal.

On the lattice the orthogonal group $O(4)$ of Euclidean space-time is 
reduced to the hypercubic group $H(4)\subset O(4)$. Consequently, for any lattice 
spacing, the traces $A_L$ and $B_L$ [\Eq{eq:ALBL}] are symmetric 
under $H(4)$ transformations and hence functions of the hypercubic invariants
$p^2$, $p^{[4]}$, $p^{[6]},p^{[8]},\ldots$ with 
\begin{equation}
  p^{[2i]} \mathrel{\mathop:}= \sum_\mu p^{2i}_\mu \qquad\forall i=2,3,4,\ldots\;\text{.}
\end{equation}
In four dimensions the first four invariants are sufficient. All remaining 
invariants, and any combination thereof, are functions of those four.\footnote{For instance
\begin{displaymath}
 p^{[10]}= \frac{5p^2}{4}p^{[8]} -\frac{5p^4}{6}p^{[6]}  + 5p^{[4]}\left( \frac{p^{[6]}}{6}+ \frac{p^6}{12} - \frac{p^2p^{[4]}}{8}\right) - \frac{p^{10}}{24}\,.
\end{displaymath}}
In the continuum limit, the renormalized traces $A(\mu^2,p^2)$ and $B(\mu^2,p^2)$ are functions of $p^2$ alone. Therefore, we can assume that to leading order in the lattice spacing $a$ the 
lattice quark propagator traces $F_L=\{A_L,B_L\}$ are of the form \cite{deSoto:2007ht}
\begin{widetext}
\begin{equation}
 \label{eq:H4expansion}
 F_L(a,p) \equiv F_L(a^2,p^2,p^{[4]},p^{[6]},p^{[8]})~\simeq~ F(a^2,p^2) 
       + a^4 p^{[4]} f_4(a^2,p^2) + a^6 p^{[6]} f_6(a^2,p^2) + a^8 p^{[8]} f_8(a^2,p^2) + \ldots \,.
\end{equation}
\end{widetext}
$F(a^2,p^2)$ contains all $O(4)$-symmetric terms, i.e., terms which are functions of $a^2p^2$ and 
$a^2$ only. The ``hypercubic'' terms describe the leading deviation from $O(4)$ symmetry.

Such an expression (up to the second term) is for example obtained from an 
$O(a^2)$ expansion of the 1-loop Wilson quark propagator (see Eq.(4.1) 
in \cite{Constantinou:2009tr}). There, $F(a^2,p^2)$ is the sum of the usual constants, 
the $\log(a^2p^2)$ term as well as the scaling violations proportional to $a^2p^2$. 
The leading hypercubic correction to $A_L$ reads $a^4 p^{[4]} f_4(a^2,p^2)$ with
\begin{equation}
 \label{eq:f1l4}
 f^{1l}_4(a^2,p^2)=\frac{c_0(g^2)+c_1(g^2)\log(a^2p^2)}{a^2p^2}
\end{equation}
where the $c_i(g^2)$'s are functions of the coupling. The log-term in 
$F(a^2,p^2)$ is multiplicatively removed by the respective renormalization 
constant, while the $a^2p^{2}$ and $a^2p^{[4]}/p^2$ terms vanish in the 
continuum limit. For any finite $a$ both terms add to the scaling
violations, but those due to the hypercubic terms also depend on the momentum 
direction: they are largest for on-axis momenta and smallest (but non-zero)
for cylinder-cut momenta (see again \Fig{fig:M_uncorr_b5p29kp13640_64x64}). 
The H4 method attempts to remove exactly those contributions to the scaling violations.

Our implementation of the H4 method is a modified version of the local 
H4 method described in \cite{deSoto:2007ht}. There, 
$f_4=c_0 + c_1/p^2 + c_2p^2$ and the constants $c_i$ are 
obtained from fits to the data for a range of $p^2$. Our fits are performed 
for individual $p^2$, but we allow coefficients to depend on $p^2$. That is, we 
do not fix the form of $f_4$ and instead write
\begin{equation}
 f_4(a^2,p^2) = \frac{c(a^2,p^2)}{a^2p^2} \,.
  \label{eq:ansatzf4}
\end{equation}
Given that our lattice propagator agrees with the continuum expression
to order $\mathcal{O}(a)$, we restrict the expansion to $\mathcal{O}(a^2)$.\footnote{Note
that small $O(a)$ corrections could still be present, 
because we use the tl-values for the correction coefficients $b_q$ and $c_q$.}
We do check, however, whether the fits improve if higher hypercubic terms are 
included. We will also analyze the functional form of $c(a^2,p^2)$.

The H4 method cannot completely remove the lattice artifacts, in particular not
the scaling violations in $F(a^2,p^2)$. However, our H4 extrapolation is performed on 
the data after removing the tree-level artifacts as described above. This 
tree-level correction already drastically reduces the scaling violations in $F(a^2,p^2)$ and 
the hypercubic terms. A subsequent application of the H4 
method further reduces the hypercubic part. The bending of the 
quark dressing at large $a^2p^2$ should flatten for instance.

For the (tree-level corrected) form factors, $A$ and $B$, and the quark wave 
function $Z=1/A$, we expect similar hypercubic expansions to hold. If this is 
the case then the leading hypercubic corrections for the mass function $M=B/A$ 
should be comparably smaller, in particular if those of $A$ and $B$ are of 
similar size. We thus expect that higher hypercubic terms dominate the 
behavior at large $p^{[2i]}$. In the continuum limit $M$ is 
renormalization-group invariant and so it is also plausible that it may 
have smaller discretization effects. 

\begin{figure*}
   \centering
   \includegraphics*[width=8.5cm]{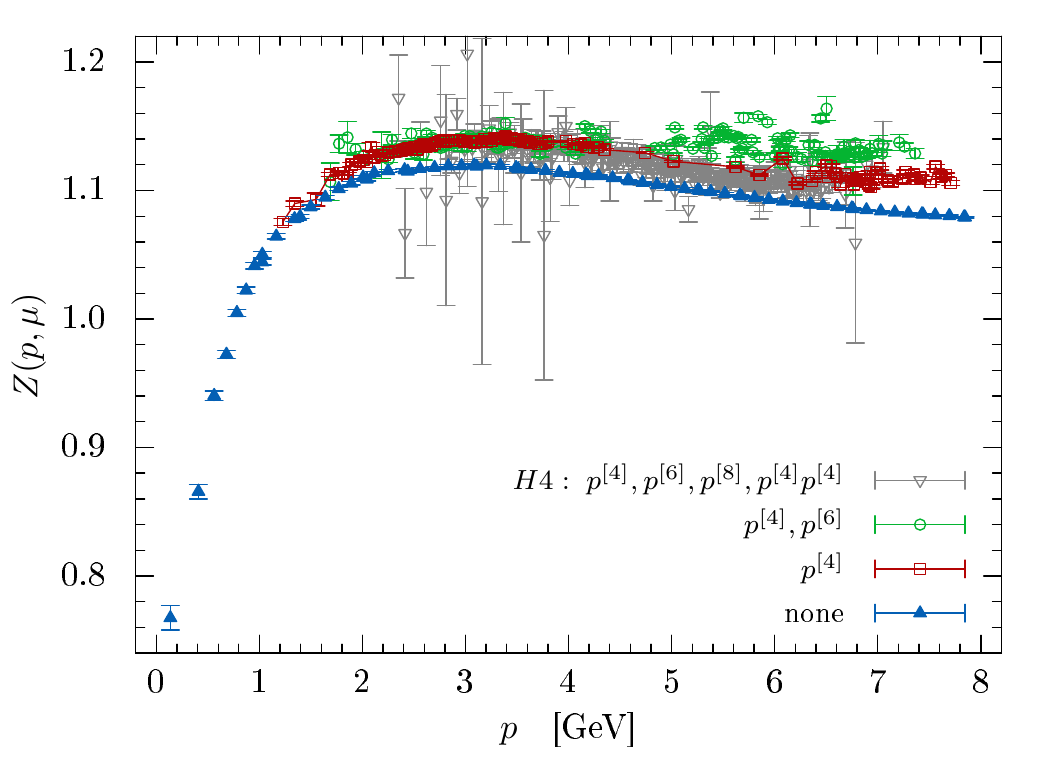} \quad
   \includegraphics*[width=8.5cm]{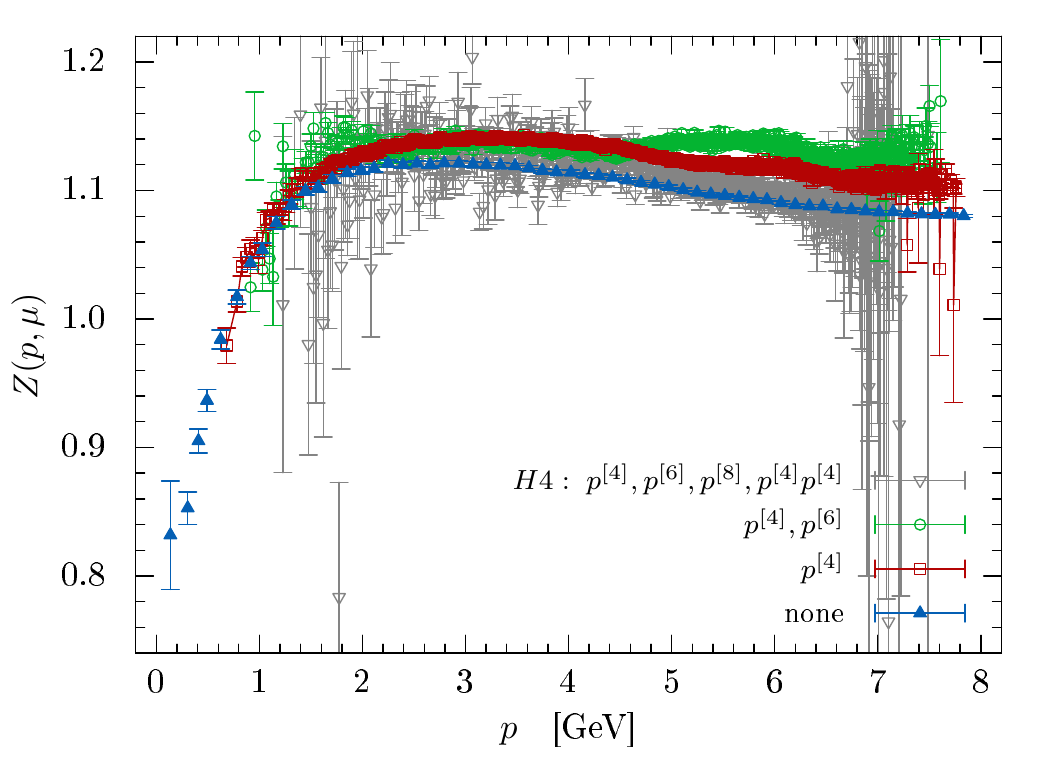} \\
   \includegraphics*[width=8.5cm]{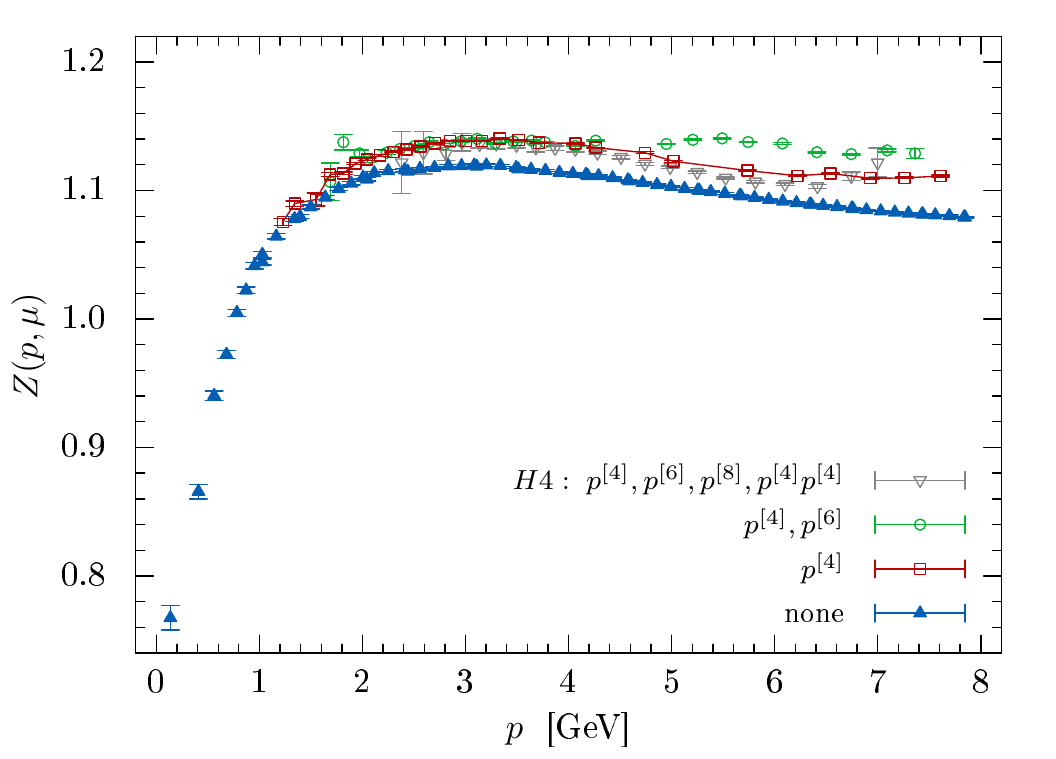} \quad
   \includegraphics*[width=8.5cm]{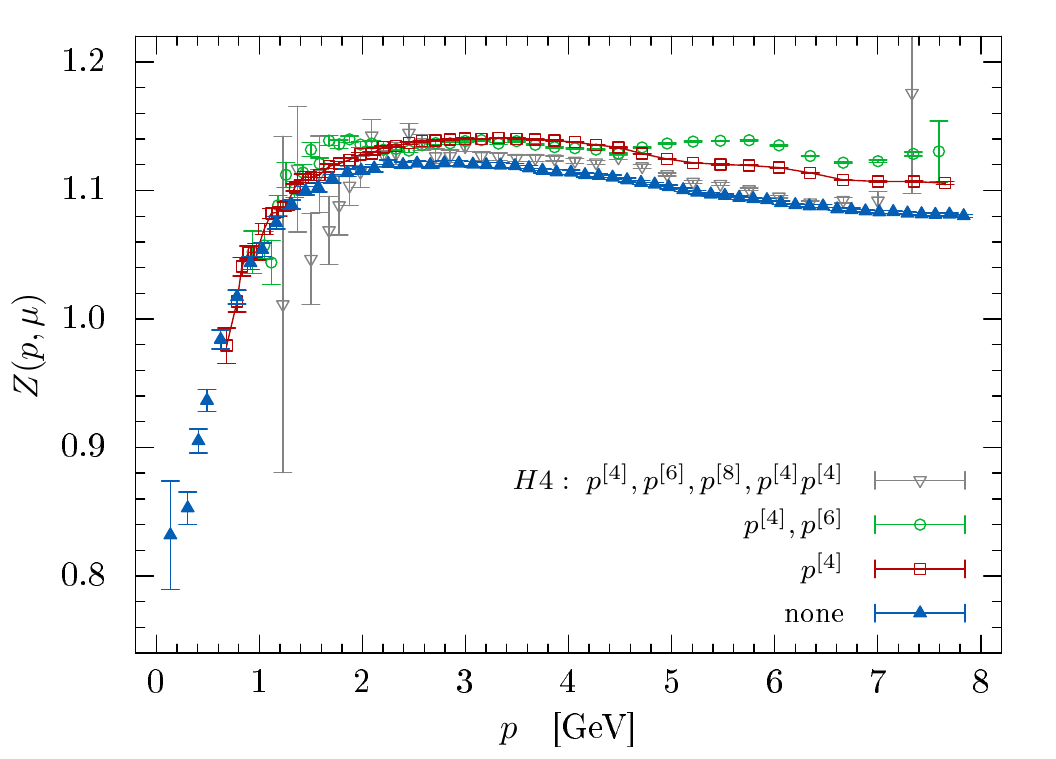}
   \caption{H4-extrapolated quark wave function for $\beta = 5.29,
     \kappa=0.13632, V=32^3\times64$ (left panels) and for
     $\beta=5.29, \kappa=0.13640, V=64^4$ (right panels).  The top
     panels show the raw data before smoothing, while the bottom
     panels show the results after averaging with a momentum
     resolution $\epsilon=0.05$ (see text for details). In the legend we list 
     the hypercubic terms included for the fit.} 
   \label{fig:bareZext}
\end{figure*}

In the next subsection we discuss the results of the H4 method applied to the 
quark wave function $Z$ and the quark mass function $M$. For the fits we group 
the lattice data $D_F$ for $F=\{Z,M\}$ wrt.\ the value of $p^2$. The number
of data points for each $p^2_i$ varies, hence the statistical error of each 
fit will vary, too. Our fit parameters are $F(a^2,p^2)$ and 
$c(a^2,p^2)$. If higher terms are included in the fit, there is an
additional parameter $c_i(a^2,p^2)$ for each of the terms $p^{[6]}$, $p^{[8]}$ 
and $p^{[4]}p^{[4]}$. The quality of a fit is monitored by the $\chi^2$-function:
\begin{multline}
    \chi^2 ( p^2_i ) = \\
    \sum_j \frac{\left[F_L(a^2,p^2_i, p^{[4]}_j, \ldots) - D_{F}(a^2,p^2_i, p^{[4]}_j, \ldots) \right]^2 }
                               {\sigma^2_{F}(p^2_i, p^{[4]}_j \ldots)}  \,,
\end{multline}
where $D_{F}$ denotes the tree-level corrected data for $F=Z$ and $M$, and 
$F_L(a^2,p^2_i, p^{[4]}_j, \ldots)$ denotes the H4 expansion in \Eq{eq:H4expansion}
with $f_4(a^2,p^2)$ in \Eq{eq:ansatzf4}. The minimization of $\chi^2$ is 
translated into finding the solution of a linear system of equations for 
$c(a^2,p^2)$, which is solved by Gauss-Jordan elimination. Statistical errors are 
estimated with the bootstrap method with a 67.5\% confidence level. The number 
of bootstrap samples is ten times the number of configurations. Fits with 
$\chi^2/d.o.f. \geqslant2$ are disregarded.

In the following, for the numerical procedure to measure $Z(p^2)$ and
$M(p^2)$ we will always assume an exact H4 hypercubic symmetry group,
which holds only for $L^4$ lattices. The results reported here will
also include the asymmetric lattice $32^3 \times 64$, but given the
volume and lattice spacing used in the simulation, we expect the
corrections due to the asymmetry to be small --- see the analysis and
the discussion in Ref.~\cite{Blossier:2010vt}. 

In \Fig{fig:ZMp2H4_b5p29kp13632_kp13640} we plot the quark wave function and 
running mass for various $a^2 p^2$ values as a function of
$a^4 p^{[4]}$ for the simulations performed with $\beta = 5.29$ on the
$32^4 \times 64$ lattice with $\kappa = 0.13632$, corresponding to
$m_\pi = 295$ MeV (upper panels) and on the $64^4$ lattice with $\kappa
= 0.13640$, corresponding to $m_\pi = 150$ MeV (lower panels). The lattice
data show a smooth behavior as a function of $a^4 p^{[4]}$, with the data for $Z(p^2)$ suggesting an essentially linear function of $a^4 p^{[4]}$, 
while the data for $a M(p^2)$ show clear deviations from a linear
behavior in $a^4 p^{[4]}$. 
From the point of view of the H4 method, the observed smooth behavior
in the various plots is quite encouraging, suggesting that it is possible
to achieve a reliable extrapolation to the O(4) symmetric limit.

\begin{figure*}
   \centering
   \mbox{\includegraphics[width=0.5\textwidth]{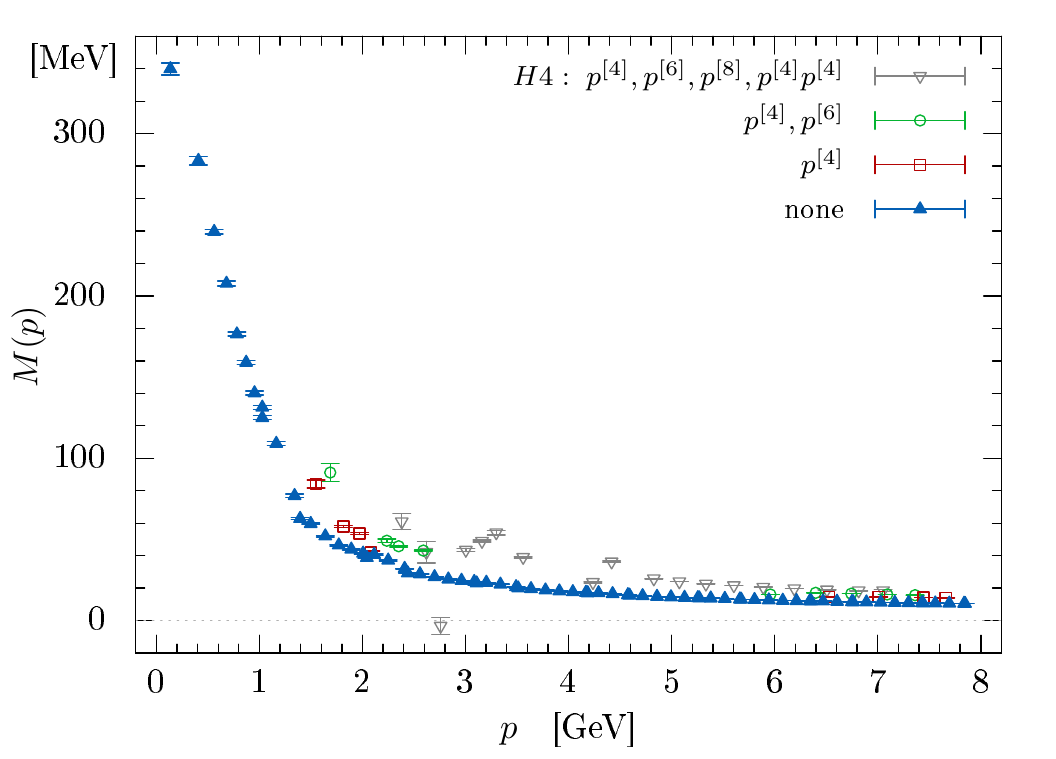}
    \includegraphics[width=0.5\textwidth]{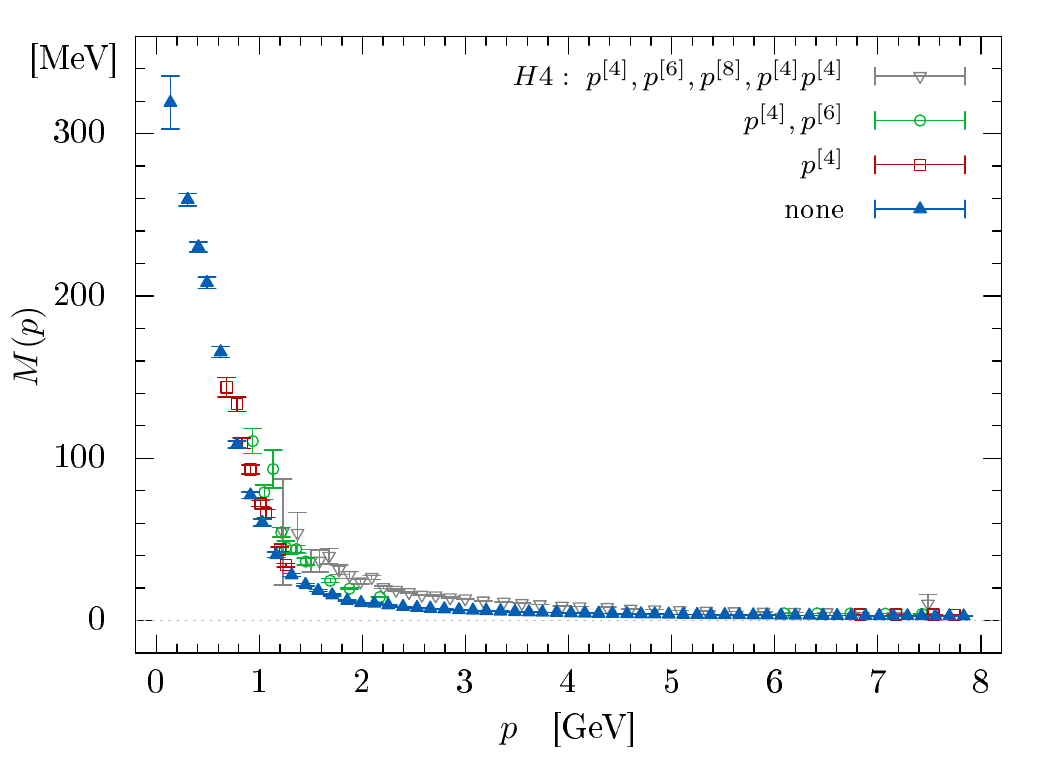}}
   \caption{H4-extrapolated running mass function $M(p^2)$ after binning 
     with a momentum resolution $\epsilon=0.05$ (see text for
     details), for $\beta = 5.29,
     \kappa=0.13632, V=32^3\times64$ (left) and
     $\beta=5.29, \kappa=0.13640, V=64^4$ (right)
     As in \Fig{fig:bareZext} we list the hypercubic terms included for the fit.}
   \label{fig:bareMext}
\end{figure*}

\subsection{Tree-level and H4-corrected data}
\label{SubSec:NumExt}

The top panels of \Fig{fig:bareZext} compare the H4-extrapolated data (open symbols)
for the quark wave function with the tree-level corrected data (full triangle)
of \Fig{fig:Z-volM-vol}. We focus again on the ensembles III and V, but the 
comparison looks similar for the other ensembles. The H4-extrapolated points result 
from three types of extrapolation: The $p^{[4]}$ points are from extrapolations 
where the hypercubic corrections are described by $f_4(a^2,p^2)$ in 
\Eq{eq:ansatzf4} alone. Open circles are from extrapolations where a 
$p^{[6]}$-correction term was included as well. The open triangles are from 
fits where also terms proportional to $p^{[8]}$ and $p^{[4]}p^{[4]}$ were included. 
All three extrapolations agree within errors up to $p\simeq4\,\text{GeV}$ ($pa \simeq 1.5$),
but the errors drastically increase when more hypercubic correction terms are included. Note 
again that points from extrapolations with $\chi^2/d.o.f. \geqslant2$ have been 
discarded and hence do not appear in \Fig{fig:bareZext}.
 
\begin{figure}
   \centering
    \includegraphics[width=0.48\textwidth]{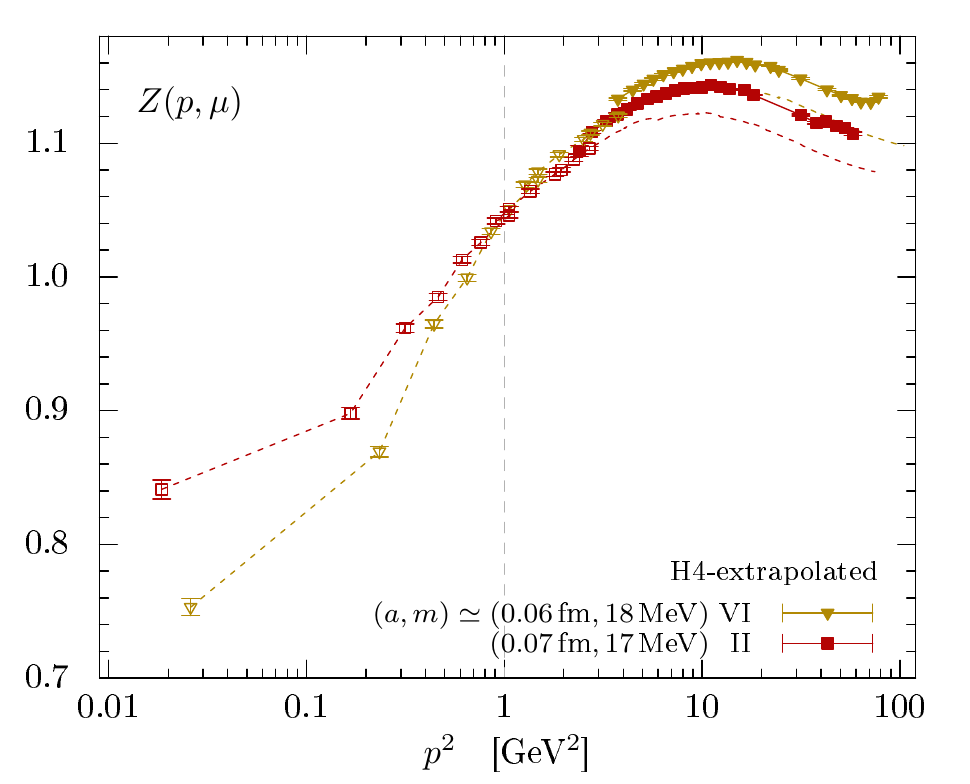}\\
    \includegraphics[width=0.48\textwidth]{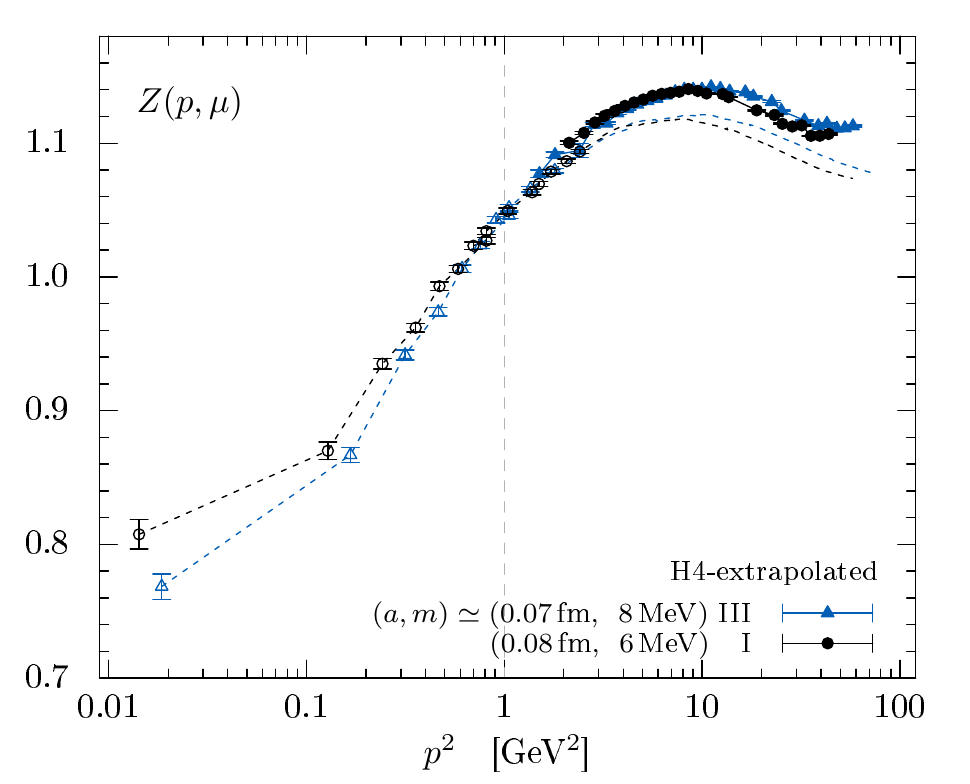}\\
    \includegraphics[width=0.48\textwidth]{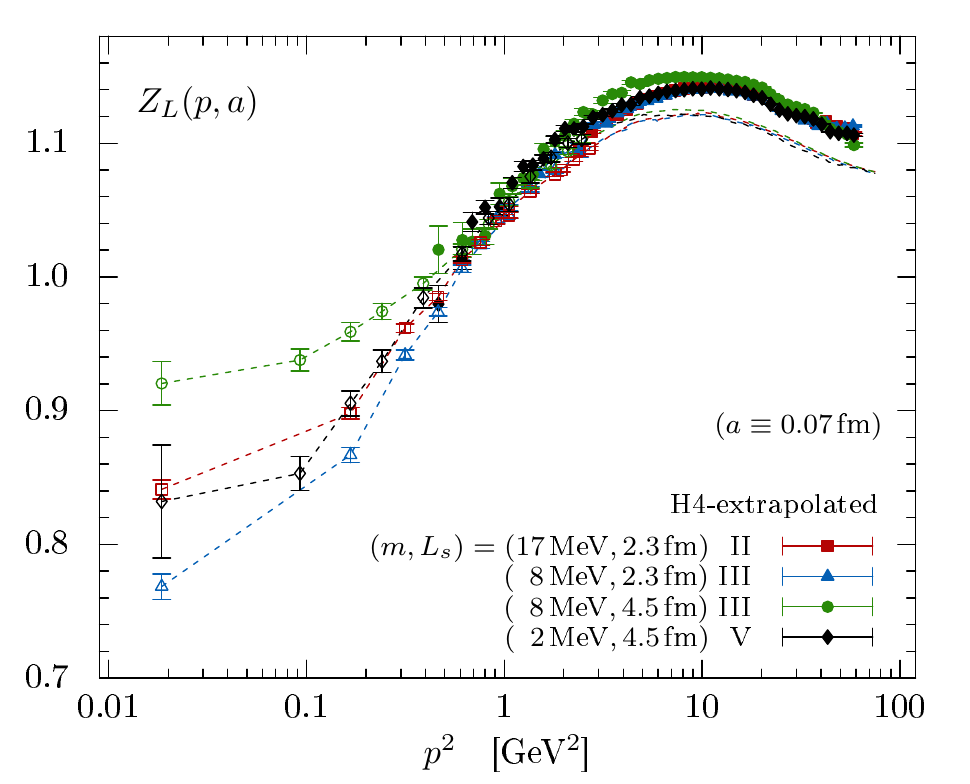}
    \caption{Quark wave function versus $p^2$ after tree-level correction (open) and H4 extrapolation (full symbols). Top and middle panel show data for different lattice spacings, $a$, but approximately same quark mass, $m=6\ldots8\,\text{MeV}$ and $m=17\ldots18\,\text{MeV}$, respectively. The bottom panel shows data for $a\equiv 0.07\,\text{fm}$ but varying $m$ and volume.}
   \label{fig:Zfinal}
\end{figure}
The bottom panels of \Fig{fig:bareZext} show the same comparison as the top but 
there the points are weighted averages of data from nearby momenta, with 
weights given by the inverse statistical error. The data binning reduces the 
statistical fluctuations drastically. We have tried different bin 
sizes by varying the momentum resolution
\begin{equation}
    \epsilon = \frac{| p' - p |}{p}
\end{equation}
and find that $\epsilon=0.05$ is a reasonable 
compromise between acceptable uncertainties, a smooth curve and a sufficient
number of data points. From the binned data we see that the three types of
H4 extrapolations give slightly different results for $p>4\,\text{GeV}$ ($pa>1.5$). 
We also find that the $p^{[6]}$ term tends to destabilize the fit, yielding an
erratic behavior at high momenta. The $64^4$ ensemble (V) has smaller 
statistics as seen in the top panels, but since there is a larger number of 
$p^2$ invariant momentum combinations within each momentum bin, after binning 
we obtain comparable results.

The H4 extrapolation works well for the quark wave function. For the running 
quark mass, however, the extrapolations perform much worse --- see Fig.~\ref{fig:bareMext}, where the H4-extrapolated $M(p)$ is shown for the same ensembles as for $Z(p)$ above. In fact, the number of fits with $\chi^2/d.o.f.\le 2$ is significantly smaller for $M(p)$, than it is for 
$Z(p^2)$. Only by including all hypercubic terms up to the $p^{[8]}$ 
and $p^{[4]}p^{[4]}$ terms can reasonable extrapolations be found. 

In \Fig{fig:bareMext} we compare the H4-extrapolated with the tree-level 
corrected data for $M(p^2)$, again by showing weighted averages of data
from nearby momenta ($\epsilon=0.05$). For large $p$, our H4-extrapolation
changes the momentum behavior of the tree-level corrected $M(p)$ only slightly, 
while for smaller $p$ the extrapolated values differ more significantly, in 
particular between $p=1.3\,\text{GeV}$ and $3\,\text{GeV}$. From the nature of 
hypercubic artifacts we would expect the opposite trend. Furthermore, the points
from the three types of extrapolations do not coincide at small $p$, while they 
tend to converge onto a single curve with the tree-level corrected (cylinder-cut)
data for high $p$. We conclude that our H4-extrapolation fails for $M$ and 
consider the tree-level corrected data in \Fig{fig:Z-volM-vol} and \ref{fig:Z-allM-all} 
as our final data for $M$.

Our results for $Z$ after tree-level correction and H4 extrapolation ($\epsilon=0.05$) are shown in \Fig{fig:Zfinal} (full symbols). If the linear H4 extrapolation in $a^2p^{[4]}/p^2$ was not successful (i.e.\ $\chi^2/d.o.f.\ge 2$), the tree-level corrected data are shown instead (open symbols). They are the same as in  \Fig{fig:Z-volM-vol} and \ref{fig:Z-allM-all}. To guide the eye and to demonstrate the shift due the H4 extrapolation we have added dashed lines connecting the tree-level corrected points. These points are not shown above $p^2\simeq3\,\text{GeV}^2$ in \Fig{fig:Zfinal} to improve the visibility of the shift. The points in the two upper panels of \Fig{fig:Zfinal} have been renormalized relative to the $a=0.07\,\text{fm}$ data at $\mu=1\,\text{GeV}$. This allows for a better comparison with the bottom panel showing unrenormalized data at fixed lattice spacing ($a=0.07\,\text{fm}$) but different volumes and quark masses.

In \Fig{fig:Zfinal} we see that the H4 extrapolation causes an upward shift of all data points above $p^2\simeq2\,\text{GeV}^2$. For the heavier quark mass sets (top panel) the H4 extrapolation causes also a slight reduction of the vertical difference for $3\,\text{GeV}^2<p^2<10\,\text{GeV}^2$. For the lighter quark mass this difference is already negligible after tree-level correction. For the single-$a$ data in the bottom panel of \Fig{fig:Zfinal} we observe that the points above $p^2=1\,\text{GeV}^2$ tend to overlap less after H4 extrapolation. This might be due to the different volume sizes which influences the quality of the H4 extrapolation there.

\subsection{Looking at the H4 expansion}

\begin{figure}
\includegraphics[width=0.48\textwidth]{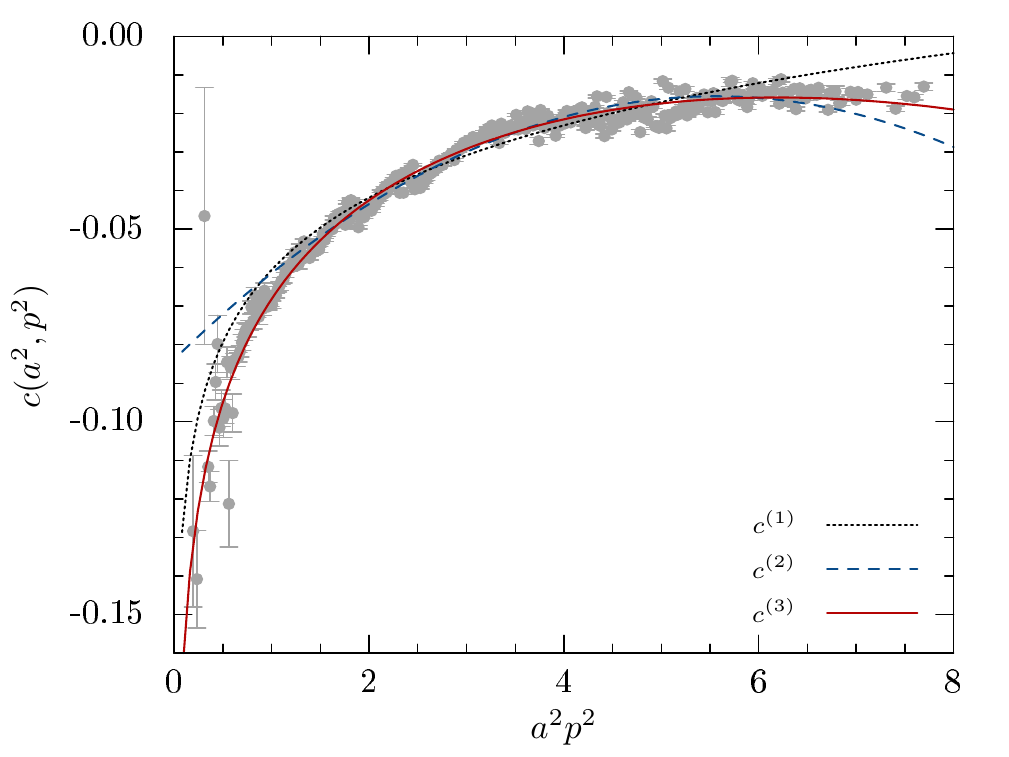}
 \caption{The expansion coefficient, $c(a^2,p^2)$, for the H4 term
   $a^2p^{[4]}/p^2$  in the expansion \eqref{eq:H4expansion} and
     \eqref{eq:ansatzf4} of $Z(p)$.  The data are for ensemble
     IV. The three lines refer to different fit ansatzes for the $a^2p^2$ dependence (see \Eq{eq:fitc}.)}
 \label{fig:c_b5p29kp13632_64x64}
\end{figure}

We close the section with a discussion on $c(a^2,p^2)$ [see
  \Eq{eq:ansatzf4}]. Remember that in our H4 extrapolation the
functional form of $c(a^2,p^2)$ is not fixed, but left as a free,
momentum dependent parameter. Hence our fit results may be
useful for future studies, e.g., when applying
global H4 extrapolations. In \Fig{fig:c_b5p29kp13632_64x64} we show
$c(a^2,p^2)$ for $Z(p)$ from ensemble IV (gray circles), together with different regression curves: 
\begin{subequations}
 \label{eq:fitc}
\begin{align}
 c^{(1)}(a^2,p^2) &= c_0 + c_1\log(a^2p^2) \,, \\
 c^{(2)}(a^2,p^2) &= c_0 + c_1 a^2p^2 + c_2 a^4p^{[4]} \,, \\
 c^{(3)}(a^2,p^2) &= c_0 + c_1\log(a^2p^2)  + c_2 a^4p^4 \,.
\end{align}
\end{subequations}
$c^{(1)}$ resembles the prefactor of the $a^2p^{[4]}/p^2$ correction term to the Wilson quark wave function in 1-loop lattice perturbation theory at $O(a^2)$ \cite[Eq.(4.1)]{Constantinou:2009tr}. The ansatz for $c^{(2)}$ is from \cite{deSoto:2007ht} and $c^{(3)}$ equals $c^{(1)}$ at small $a^2p^2$ but includes an additional $a^4p^4$ term to describe the bending for $a^2p^2>4$. We see in \Fig{fig:c_b5p29kp13632_64x64} that $c^{(3)}$ gives a good description of our fit results for $c(a^2,p^2)$, while the other two curves give a poor description if fitted to the whole $a^2p^2$ range. If the fit range for $c^{(1)}$ is restricted to $a^2p^2<4$, $c^{(1)}$ coincides with $c^{(3)}$ up to $a^2p^2=3.5$. Putting the same constraint for $c^{(2)}$, we find that $c^{(2)}$ and $c^{(3)}$ overlap for $1<a^2p^2<4$.

In short, our analysis of the fitted H4 expansion seems to favor the functional form 
$c^{(3)}(a^2,p^2)$ which we adapted from 1-loop lattice perturbation theory. It better
reproduces the observed behavior for large $a^2p^2$ and hence could improve
global H4 fits, e.g., as performed in \cite{deSoto:2007ht}.

\section{Summary}
\label{sec:summary}

We have studied the quark propagator in Landau gauge on $N_f=2$ gauge field configurations using $\order(a)$-improved Wilson fermions for both large and almost physical quark masses.

In agreement with previous studies, we find that the quark wave function, $Z(p^2)$, is infrared suppressed and the quark mass function, $M(p^2)$, shows the same qualitative features as in previous studies with other discretizations, with a dynamically quark mass developing for momenta below 1--2 GeV, tending to a value $M(0)\approx300$ MeV in the chiral limit. Compared to results using staggered and overlap fermions, $M(p^2)$ drops more quickly when increasing $p$ from $200\,\text{MeV}$ to $p=1\,\text{GeV}$. We also do not see a clear sign of a plateau at small momenta. Lattice data below $p=200\,\text{MeV}$ are needed to determine whether such a plateau exists.

Our final lattice data for $M(p)$ are shown in the right panels of Figs.\,\ref{fig:Z-volM-vol} and \ref{fig:Z-allM-all}. These data were obtained applying the hybrid tree-level correction described above and restricting to cylinder-cut momenta. We tried to reduce the remaining hypercubic artifacts using the H4 method, but found that a linear extrapolation in $a^2p^{[4]}/p^2$ at fixed $p^2$ fails for $M(p)$. Higher hypercubic corrections terms are needed to reach at reasonable $\chi^2/d.o.f$ values. However, this introduces rather large uncertainties in the extrapolation, in particular a systematic error: at lower $ap$ ($ap=3\ldots8$) the corrections come out to be significantly larger than at higher momenta ($ap>16$), where they are almost negligible. For $ap<3$ the H4 extrapolations gives no reasonable $\chi^2/d.o.f.$ values.

Our final results for $Z(p^2)$ are shown in \Fig{fig:Zfinal}. There we
show the results for cylinder-cut momenta after hybrid tree-level
correction and our linear H4 extrapolation. This extrapolation has
been successful for momenta above $p\simeq 1\,\text{GeV}$ and shifted the point where $Z$ starts to bend down to a larger $p^2$. Lattice spacing effects could not completely be eliminated but much reduced by applying those techniques. For large momenta, the wave function is essentially independent of quark mass and volume, while the infrared suppression at small momentum becomes stronger for smaller quark mass. We also find competing finite volume and lattice spacing effects at small $p$: the suppression becomes weaker with larger volumes, but stronger towards smaller lattice spacings. 

Our study is the first to use fully dynamical $\order(a)$-improved Wilson fermions to access the quark wave and mass function. Lattice calculations, for example, of the nonperturbative RI'(S)MOM renormalization constants for hadron physics (see, e.g., \cite{Gockeler:2010yr}), typically did not use those. We were able to reduce lattice spacing artifacts to percent level and at the same time studied a range of quark masses down to an almost physical value. 

\begin{acknowledgments}
We thank the RQCD collaboration for giving us access to their gauge configurations. The gauge fixing and calculations of the fermion propagators were performed on the HLRN supercomputing facilities in Berlin and Hanover, as part of the project bep00046 by Michael M\"uller-Preu{\ss}ker to whose memory this paper is dedicated. He was a member of the collaboration and is sorely missed. JIS acknowledges the support and hospitality of the CSSM, where part of this work was carried out. JIS has been supported by Science Foundation Ireland grant 11/RFP.1/PHY/1362. AS acknowledges support by the BMBF under grant No.\ 05P15SJFAA (FAIR-APPA-SPARC) and by the DFG Research Training Group GRK1523. OO acknowledges support from FAPESP Grant Number \mbox{2017/01142-4}. PJS acknowledges support by FCT under contracts SFRH/BPD/40998/2007 and SFRH/BPD/109971/2015. 
\end{acknowledgments}

\begin{appendix}

\section{On the Tree-Level Corrected Running Quark Mass}
\label{sec:multvhyb}

\begin{figure}[t]
\includegraphics[width=\linewidth]{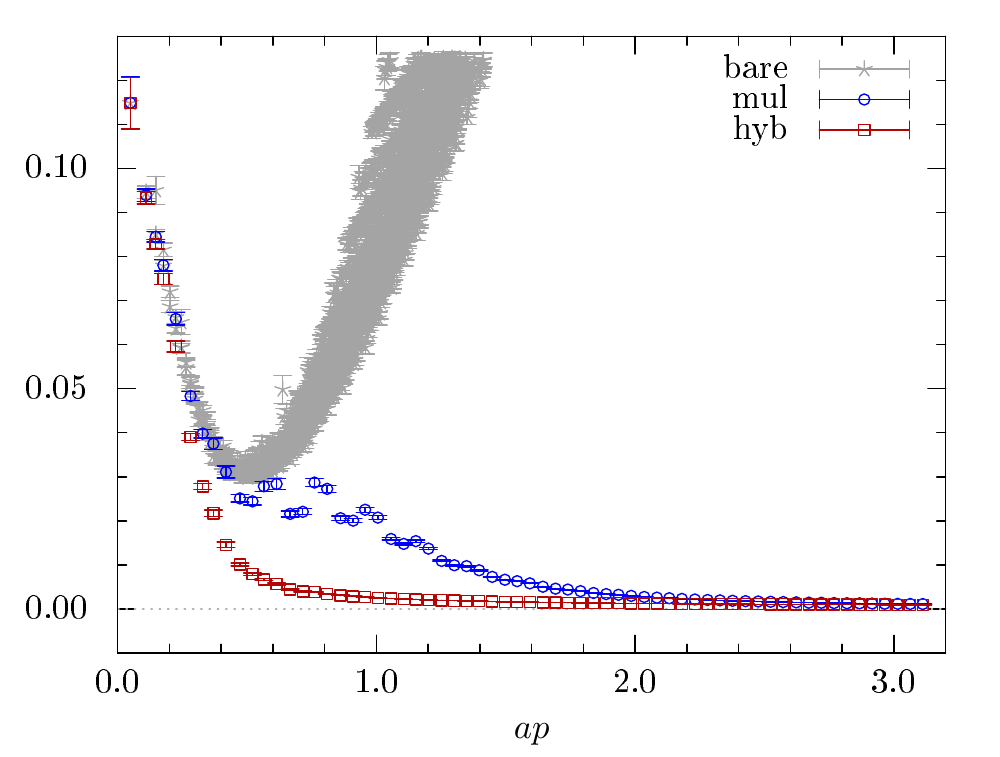}
\caption{Hybrid and multiplicatively tree-level corrected mass function $aM(p)$
  versus the uncorrected data for ensemble V as a function of $ap$.  
  The momenta for the corrected data have been cylinder cut with a radius of 1 
  lattice momentum unit. The uncorrected data have not been cut.}
\label{fig:M-hybmult}
\end{figure}

For the tree-level correction of the running quark mass data we used the hybrid
prescription described in Section \ref{sec:corrections} and \cite{Skullerud:2001aw}. 
We chose this because we found that it generally provides a smoother momentum dependence 
for $M$ than, for example, the multiplicative tree-level correction (see Section 
\ref{sec:corrections} for a definition). To demonstrate the advantage of the hybrid
correction we compare in \Fig{fig:M-hybmult} tree-level corrected data for
the two prescriptions. We chose the lightest quark mass ensemble V for this. 
We see hat both corrections give comparable results (within one standard deviation) 
at low momenta ($ap<0.3$), but the multiplicatively corrected points deviate 
strongly from the hybrid corrected points in the mid-momentum regime ($0.3<ap<2$). 
At large $ap$ the two curves seem to approach each other again, but still at 
$p = 8.6$ GeV, the highest momenta accessible in our study, the two $M$'s differ
by many standard deviations: for the multiplicatively corrected quark mass 
function we have $M(8.6\,\text{GeV}) = 3.1176 \pm 0.0019\,\text{MeV}$, while for 
the hybrid corrected mass function we have 
$M(8.6\,\text{GeV}) = 2.602 \pm 0.0094\,\text{MeV}$.

A more detailed investigation of the multiplicatively corrected mass function in 
the mid-momentum regime shows that it is the lattice momenta along the diagonal 
which deviate most strongly from the general trend in the data. This somewhat 
surprising result can  be understood looking again at \Fig{fig:Z0M0}: For 
diagonal momenta, the mass function 
$aM_{\mathrm{rot}}^{(0)}(p)=am + a\Delta M_{\mathrm{rot}}^{(0)}(p)$
is very close to $am$ but stays below $am$ within $0\le a^2p^2 \le 4$. 
For on-axis momenta, on the other hand $a M_{\mathrm{rot}}^{(0)}(p)$ 
rises fast with $a^2p^2$. The smallness of $aM_{\mathrm{rot}}^{(0)}(p)$ 
for diagonal momenta $0\le a^2p^2 \le 4$ results in
small values for $Z_m=aM_{\mathrm{rot}}^{(0)}(p)/am$. Applying $Z_m$ 
artificially enhances the multiplicatively corrected $M = M_L/Z_m$, which may 
even become negative for very small $am$. The hybrid correction does not have 
this feature, and provides a smooth curve for $M$ for all momenta.

\end{appendix}

\bibliography{references}

\begin{thebibliography}{40}%
\makeatletter
\providecommand \@ifxundefined [1]{%
 \@ifx{#1\undefined}
}%
\providecommand \@ifnum [1]{%
 \ifnum #1\expandafter \@firstoftwo
 \else \expandafter \@secondoftwo
 \fi
}%
\providecommand \@ifx [1]{%
 \ifx #1\expandafter \@firstoftwo
 \else \expandafter \@secondoftwo
 \fi
}%
\providecommand \natexlab [1]{#1}%
\providecommand \enquote  [1]{``#1''}%
\providecommand \bibnamefont  [1]{#1}%
\providecommand \bibfnamefont [1]{#1}%
\providecommand \citenamefont [1]{#1}%
\providecommand \href@noop [0]{\@secondoftwo}%
\providecommand \href [0]{\begingroup \@sanitize@url \@href}%
\providecommand \@href[1]{\@@startlink{#1}\@@href}%
\providecommand \@@href[1]{\endgroup#1\@@endlink}%
\providecommand \@sanitize@url [0]{\catcode `\\12\catcode `\$12\catcode
  `\&12\catcode `\#12\catcode `\^12\catcode `\_12\catcode `\%12\relax}%
\providecommand \@@startlink[1]{}%
\providecommand \@@endlink[0]{}%
\providecommand \url  [0]{\begingroup\@sanitize@url \@url }%
\providecommand \@url [1]{\endgroup\@href {#1}{\urlprefix }}%
\providecommand \urlprefix  [0]{URL }%
\providecommand \Eprint [0]{\href }%
\providecommand \doibase [0]{http://dx.doi.org/}%
\providecommand \selectlanguage [0]{\@gobble}%
\providecommand \bibinfo  [0]{\@secondoftwo}%
\providecommand \bibfield  [0]{\@secondoftwo}%
\providecommand \translation [1]{[#1]}%
\providecommand \BibitemOpen [0]{}%
\providecommand \bibitemStop [0]{}%
\providecommand \bibitemNoStop [0]{.\EOS\space}%
\providecommand \EOS [0]{\spacefactor3000\relax}%
\providecommand \BibitemShut  [1]{\csname bibitem#1\endcsname}%
\let\auto@bib@innerbib\@empty
\bibitem [{\citenamefont {Roberts}\ and\ \citenamefont
  {Williams}(1994)}]{Roberts:1994dr}%
  \BibitemOpen
  \bibfield  {author} {\bibinfo {author} {\bibfnamefont {Craig~D.}\
  \bibnamefont {Roberts}}\ and\ \bibinfo {author} {\bibfnamefont {Anthony~G.}\
  \bibnamefont {Williams}},\ }\bibfield  {title} {\enquote {\bibinfo {title}
  {{Dyson-Schwinger equations and their application to hadronic physics}},}\
  }\href {\doibase 10.1016/0146-6410(94)90049-3} {\bibfield  {journal}
  {\bibinfo  {journal} {Prog. Part. Nucl. Phys.}\ }\textbf {\bibinfo {volume}
  {33}},\ \bibinfo {pages} {477--575} (\bibinfo {year} {1994})},\ \Eprint
  {http://arxiv.org/abs/hep-ph/9403224} {arXiv:hep-ph/9403224 [hep-ph]}
  \BibitemShut {NoStop}%
\bibitem [{\citenamefont {Fischer}(2006)}]{Fischer:2006ub}%
  \BibitemOpen
  \bibfield  {author} {\bibinfo {author} {\bibfnamefont {Christian~S.}\
  \bibnamefont {Fischer}},\ }\bibfield  {title} {\enquote {\bibinfo {title}
  {{Infrared properties of QCD from Dyson-Schwinger equations}},}\ }\href
  {\doibase 10.1088/0954-3899/32/8/R02} {\bibfield  {journal} {\bibinfo
  {journal} {J. Phys.}\ }\textbf {\bibinfo {volume} {G32}},\ \bibinfo {pages}
  {R253--R291} (\bibinfo {year} {2006})},\ \Eprint
  {http://arxiv.org/abs/hep-ph/0605173} {arXiv:hep-ph/0605173 [hep-ph]}
  \BibitemShut {NoStop}%
\bibitem [{\citenamefont {Cloet}\ and\ \citenamefont
  {Roberts}(2014)}]{Cloet:2013jya}%
  \BibitemOpen
  \bibfield  {author} {\bibinfo {author} {\bibfnamefont {Ian~C.}\ \bibnamefont
  {Cloet}}\ and\ \bibinfo {author} {\bibfnamefont {Craig~D.}\ \bibnamefont
  {Roberts}},\ }\bibfield  {title} {\enquote {\bibinfo {title} {{Explanation
  and Prediction of Observables using Continuum Strong QCD}},}\ }\href
  {\doibase 10.1016/j.ppnp.2014.02.001} {\bibfield  {journal} {\bibinfo
  {journal} {Prog. Part. Nucl. Phys.}\ }\textbf {\bibinfo {volume} {77}},\
  \bibinfo {pages} {1--69} (\bibinfo {year} {2014})},\ \Eprint
  {http://arxiv.org/abs/1310.2651} {arXiv:1310.2651 [nucl-th]} \BibitemShut
  {NoStop}%
\bibitem [{\citenamefont {Eichmann}\ \emph {et~al.}(2016)\citenamefont
  {Eichmann}, \citenamefont {Sanchis-Alepuz}, \citenamefont {Williams},
  \citenamefont {Alkofer},\ and\ \citenamefont {Fischer}}]{Eichmann:2016yit}%
  \BibitemOpen
  \bibfield  {author} {\bibinfo {author} {\bibfnamefont {Gernot}\ \bibnamefont
  {Eichmann}}, \bibinfo {author} {\bibfnamefont {Helios}\ \bibnamefont
  {Sanchis-Alepuz}}, \bibinfo {author} {\bibfnamefont {Richard}\ \bibnamefont
  {Williams}}, \bibinfo {author} {\bibfnamefont {Reinhard}\ \bibnamefont
  {Alkofer}}, \ and\ \bibinfo {author} {\bibfnamefont {Christian~S.}\
  \bibnamefont {Fischer}},\ }\bibfield  {title} {\enquote {\bibinfo {title}
  {{Baryons as relativistic three-quark bound states}},}\ }\href {\doibase
  10.1016/j.ppnp.2016.07.001} {\bibfield  {journal} {\bibinfo  {journal} {Prog.
  Part. Nucl. Phys.}\ }\textbf {\bibinfo {volume} {91}},\ \bibinfo {pages}
  {1--100} (\bibinfo {year} {2016})},\ \Eprint
  {http://arxiv.org/abs/1606.09602} {arXiv:1606.09602 [hep-ph]} \BibitemShut
  {NoStop}%
\bibitem [{\citenamefont {Williams}(2015)}]{Williams:2014iea}%
  \BibitemOpen
  \bibfield  {author} {\bibinfo {author} {\bibfnamefont {Richard}\ \bibnamefont
  {Williams}},\ }\bibfield  {title} {\enquote {\bibinfo {title} {{The
  quark-gluon vertex in Landau gauge bound-state studies}},}\ }\href {\doibase
  10.1140/epja/i2015-15057-4} {\bibfield  {journal} {\bibinfo  {journal} {Eur.
  Phys. J.}\ }\textbf {\bibinfo {volume} {A51}},\ \bibinfo {pages} {57}
  (\bibinfo {year} {2015})},\ \Eprint {http://arxiv.org/abs/1404.2545}
  {arXiv:1404.2545 [hep-ph]} \BibitemShut {NoStop}%
\bibitem [{\citenamefont {Williams}\ \emph {et~al.}(2016)\citenamefont
  {Williams}, \citenamefont {Fischer},\ and\ \citenamefont
  {Heupel}}]{Williams:2015cvx}%
  \BibitemOpen
  \bibfield  {author} {\bibinfo {author} {\bibfnamefont {Richard}\ \bibnamefont
  {Williams}}, \bibinfo {author} {\bibfnamefont {Christian~S.}\ \bibnamefont
  {Fischer}}, \ and\ \bibinfo {author} {\bibfnamefont {Walter}\ \bibnamefont
  {Heupel}},\ }\bibfield  {title} {\enquote {\bibinfo {title} {{Light mesons in
  QCD and unquenching effects from the 3PI effective action}},}\ }\href
  {\doibase 10.1103/PhysRevD.93.034026} {\bibfield  {journal} {\bibinfo
  {journal} {Phys. Rev.}\ }\textbf {\bibinfo {volume} {D93}},\ \bibinfo {pages}
  {034026} (\bibinfo {year} {2016})},\ \Eprint
  {http://arxiv.org/abs/1512.00455} {arXiv:1512.00455 [hep-ph]} \BibitemShut
  {NoStop}%
\bibitem [{\citenamefont {Cyrol}\ \emph {et~al.}(2018)\citenamefont {Cyrol},
  \citenamefont {Mitter}, \citenamefont {Pawlowski},\ and\ \citenamefont
  {Strodthoff}}]{Cyrol:2017ewj}%
  \BibitemOpen
  \bibfield  {author} {\bibinfo {author} {\bibfnamefont {Anton~K.}\
  \bibnamefont {Cyrol}}, \bibinfo {author} {\bibfnamefont {Mario}\ \bibnamefont
  {Mitter}}, \bibinfo {author} {\bibfnamefont {Jan~M.}\ \bibnamefont
  {Pawlowski}}, \ and\ \bibinfo {author} {\bibfnamefont {Nils}\ \bibnamefont
  {Strodthoff}},\ }\bibfield  {title} {\enquote {\bibinfo {title}
  {{Nonperturbative quark, gluon, and meson correlators of unquenched QCD}},}\
  }\href {\doibase 10.1103/PhysRevD.97.054006} {\bibfield  {journal} {\bibinfo
  {journal} {Phys. Rev.}\ }\textbf {\bibinfo {volume} {D97}},\ \bibinfo {pages}
  {054006} (\bibinfo {year} {2018})},\ \Eprint
  {http://arxiv.org/abs/1706.06326} {arXiv:1706.06326 [hep-ph]} \BibitemShut
  {NoStop}%
\bibitem [{\citenamefont {Aguilar}\ \emph {et~al.}(2018)\citenamefont
  {Aguilar}, \citenamefont {Cardona}, \citenamefont {Ferreira},\ and\
  \citenamefont {Papavassiliou}}]{Aguilar:2018epe}%
  \BibitemOpen
  \bibfield  {author} {\bibinfo {author} {\bibfnamefont {A.~C.}\ \bibnamefont
  {Aguilar}}, \bibinfo {author} {\bibfnamefont {J.~C.}\ \bibnamefont
  {Cardona}}, \bibinfo {author} {\bibfnamefont {M.~N.}\ \bibnamefont
  {Ferreira}}, \ and\ \bibinfo {author} {\bibfnamefont {J.}~\bibnamefont
  {Papavassiliou}},\ }\bibfield  {title} {\enquote {\bibinfo {title} {{Quark
  gap equation with non-abelian Ball-Chiu vertex}},}\ }\href {\doibase
  10.1103/PhysRevD.98.014002} {\bibfield  {journal} {\bibinfo  {journal} {Phys.
  Rev.}\ }\textbf {\bibinfo {volume} {D98}},\ \bibinfo {pages} {014002}
  (\bibinfo {year} {2018})},\ \Eprint {http://arxiv.org/abs/1804.04229}
  {arXiv:1804.04229 [hep-ph]} \BibitemShut {NoStop}%
\bibitem [{\citenamefont {Becirevic}\ \emph
  {et~al.}(2000{\natexlab{a}})\citenamefont {Becirevic}, \citenamefont
  {Lubicz}, \citenamefont {Martinelli},\ and\ \citenamefont
  {Testa}}]{Becirevic:1999rv}%
  \BibitemOpen
  \bibfield  {author} {\bibinfo {author} {\bibfnamefont {D.}~\bibnamefont
  {Becirevic}}, \bibinfo {author} {\bibfnamefont {V.}~\bibnamefont {Lubicz}},
  \bibinfo {author} {\bibfnamefont {G.}~\bibnamefont {Martinelli}}, \ and\
  \bibinfo {author} {\bibfnamefont {M.}~\bibnamefont {Testa}},\ }\bibfield
  {title} {\enquote {\bibinfo {title} {{Quark masses and renormalization
  constants from quark propagator and three point functions}},}\ }\href
  {\doibase 10.1016/S0920-5632(00)91828-0} {\bibfield  {journal} {\bibinfo
  {journal} {Nucl. Phys. Proc. Suppl.}\ }\textbf {\bibinfo {volume} {83}},\
  \bibinfo {pages} {863--865} (\bibinfo {year} {2000}{\natexlab{a}})},\ \Eprint
  {http://arxiv.org/abs/hep-lat/9909039} {arXiv:hep-lat/9909039 [hep-lat]}
  \BibitemShut {NoStop}%
\bibitem [{\citenamefont {Becirevic}\ \emph
  {et~al.}(2000{\natexlab{b}})\citenamefont {Becirevic}, \citenamefont
  {G{\'i}menez}, \citenamefont {Lubicz},\ and\ \citenamefont
  {Martinelli}}]{Becirevic:1999kb}%
  \BibitemOpen
  \bibfield  {author} {\bibinfo {author} {\bibfnamefont {Damir}\ \bibnamefont
  {Becirevic}}, \bibinfo {author} {\bibfnamefont {Vicente}\ \bibnamefont
  {G{\'i}menez}}, \bibinfo {author} {\bibfnamefont {Vittorio}\ \bibnamefont
  {Lubicz}}, \ and\ \bibinfo {author} {\bibfnamefont {Guido}\ \bibnamefont
  {Martinelli}},\ }\bibfield  {title} {\enquote {\bibinfo {title} {{Light quark
  masses from lattice quark propagators at large momenta}},}\ }\href {\doibase
  10.1103/PhysRevD.61.114507} {\bibfield  {journal} {\bibinfo  {journal} {Phys.
  Rev.}\ }\textbf {\bibinfo {volume} {D61}},\ \bibinfo {pages} {114507}
  (\bibinfo {year} {2000}{\natexlab{b}})},\ \Eprint
  {http://arxiv.org/abs/hep-lat/9909082} {arXiv:hep-lat/9909082 [hep-lat]}
  \BibitemShut {NoStop}%
\bibitem [{\citenamefont {Skullerud}\ and\ \citenamefont
  {Williams}(2001)}]{Skullerud:2000un}%
  \BibitemOpen
  \bibfield  {author} {\bibinfo {author} {\bibfnamefont {Jon~Ivar}\
  \bibnamefont {Skullerud}}\ and\ \bibinfo {author} {\bibfnamefont
  {Anthony~G.}\ \bibnamefont {Williams}},\ }\bibfield  {title} {\enquote
  {\bibinfo {title} {{Quark propagator in Landau gauge}},}\ }\href {\doibase
  10.1103/PhysRevD.63.054508} {\bibfield  {journal} {\bibinfo  {journal} {Phys.
  Rev.}\ }\textbf {\bibinfo {volume} {D63}},\ \bibinfo {pages} {054508}
  (\bibinfo {year} {2001})},\ \Eprint {http://arxiv.org/abs/hep-lat/0007028}
  {arXiv:hep-lat/0007028 [hep-lat]} \BibitemShut {NoStop}%
\bibitem [{\citenamefont {Skullerud}\ \emph {et~al.}(2001)\citenamefont
  {Skullerud}, \citenamefont {Leinweber},\ and\ \citenamefont
  {Williams}}]{Skullerud:2001aw}%
  \BibitemOpen
  \bibfield  {author} {\bibinfo {author} {\bibfnamefont {Jonivar}\ \bibnamefont
  {Skullerud}}, \bibinfo {author} {\bibfnamefont {Derek~B.}\ \bibnamefont
  {Leinweber}}, \ and\ \bibinfo {author} {\bibfnamefont {Anthony~G.}\
  \bibnamefont {Williams}},\ }\bibfield  {title} {\enquote {\bibinfo {title}
  {{Nonperturbative improvement and tree level correction of the quark
  propagator}},}\ }\href {\doibase 10.1103/PhysRevD.64.074508} {\bibfield
  {journal} {\bibinfo  {journal} {Phys. Rev.}\ }\textbf {\bibinfo {volume}
  {D64}},\ \bibinfo {pages} {074508} (\bibinfo {year} {2001})},\ \Eprint
  {http://arxiv.org/abs/hep-lat/0102013} {arXiv:hep-lat/0102013 [hep-lat]}
  \BibitemShut {NoStop}%
\bibitem [{\citenamefont {Boucaud}\ \emph {et~al.}(2003)\citenamefont
  {Boucaud}, \citenamefont {de~Soto}, \citenamefont {Leroy}, \citenamefont
  {Le~Yaouanc}, \citenamefont {Micheli}, \citenamefont {Moutarde},
  \citenamefont {P{\`e}~ne},\ and\ \citenamefont
  {Rodr{\'i}guez-Quintero}}]{Boucaud:2003dx}%
  \BibitemOpen
  \bibfield  {author} {\bibinfo {author} {\bibfnamefont {Philippe}\
  \bibnamefont {Boucaud}}, \bibinfo {author} {\bibfnamefont {F.}~\bibnamefont
  {de~Soto}}, \bibinfo {author} {\bibfnamefont {J.~P.}\ \bibnamefont {Leroy}},
  \bibinfo {author} {\bibfnamefont {A.}~\bibnamefont {Le~Yaouanc}}, \bibinfo
  {author} {\bibfnamefont {J.}~\bibnamefont {Micheli}}, \bibinfo {author}
  {\bibfnamefont {H.}~\bibnamefont {Moutarde}}, \bibinfo {author}
  {\bibfnamefont {O.}~\bibnamefont {P{\`e}~ne}}, \ and\ \bibinfo {author}
  {\bibfnamefont {J.}~\bibnamefont {Rodr{\'i}guez-Quintero}},\ }\bibfield
  {title} {\enquote {\bibinfo {title} {{Quark propagator and vertex: Systematic
  corrections of hypercubic artifacts from lattice simulations}},}\ }\href
  {\doibase 10.1016/j.physletb.2003.08.065} {\bibfield  {journal} {\bibinfo
  {journal} {Phys. Lett.}\ }\textbf {\bibinfo {volume} {B575}},\ \bibinfo
  {pages} {256--267} (\bibinfo {year} {2003})},\ \Eprint
  {http://arxiv.org/abs/hep-lat/0307026} {arXiv:hep-lat/0307026 [hep-lat]}
  \BibitemShut {NoStop}%
\bibitem [{\citenamefont {Bowman}\ \emph {et~al.}(2002)\citenamefont {Bowman},
  \citenamefont {Heller},\ and\ \citenamefont {Williams}}]{Bowman:2002bm}%
  \BibitemOpen
  \bibfield  {author} {\bibinfo {author} {\bibfnamefont {Patrick~O.}\
  \bibnamefont {Bowman}}, \bibinfo {author} {\bibfnamefont {Urs~M.}\
  \bibnamefont {Heller}}, \ and\ \bibinfo {author} {\bibfnamefont {Anthony~G.}\
  \bibnamefont {Williams}},\ }\bibfield  {title} {\enquote {\bibinfo {title}
  {{Lattice quark propagator with staggered quarks in Landau and Laplacian
  gauges}},}\ }\href {\doibase 10.1103/PhysRevD.66.014505} {\bibfield
  {journal} {\bibinfo  {journal} {Phys. Rev.}\ }\textbf {\bibinfo {volume}
  {D66}},\ \bibinfo {pages} {014505} (\bibinfo {year} {2002})},\ \Eprint
  {http://arxiv.org/abs/hep-lat/0203001} {arXiv:hep-lat/0203001 [hep-lat]}
  \BibitemShut {NoStop}%
\bibitem [{\citenamefont {Parappilly}\ \emph {et~al.}(2006)\citenamefont
  {Parappilly}, \citenamefont {Bowman}, \citenamefont {Heller}, \citenamefont
  {Leinweber}, \citenamefont {Williams},\ and\ \citenamefont
  {Zhang}}]{Parappilly:2005ei}%
  \BibitemOpen
  \bibfield  {author} {\bibinfo {author} {\bibfnamefont {Maria~B.}\
  \bibnamefont {Parappilly}}, \bibinfo {author} {\bibfnamefont {Patrick~O.}\
  \bibnamefont {Bowman}}, \bibinfo {author} {\bibfnamefont {Urs~M.}\
  \bibnamefont {Heller}}, \bibinfo {author} {\bibfnamefont {Derek~B.}\
  \bibnamefont {Leinweber}}, \bibinfo {author} {\bibfnamefont {Anthony~G.}\
  \bibnamefont {Williams}}, \ and\ \bibinfo {author} {\bibfnamefont {J.~B}\
  \bibnamefont {Zhang}},\ }\bibfield  {title} {\enquote {\bibinfo {title}
  {{Scaling behavior of quark propagator in full QCD}},}\ }\href {\doibase
  10.1103/PhysRevD.73.054504} {\bibfield  {journal} {\bibinfo  {journal} {Phys.
  Rev.}\ }\textbf {\bibinfo {volume} {D73}},\ \bibinfo {pages} {054504}
  (\bibinfo {year} {2006})},\ \Eprint {http://arxiv.org/abs/hep-lat/0511007}
  {arXiv:hep-lat/0511007 [hep-lat]} \BibitemShut {NoStop}%
\bibitem [{\citenamefont {Bowman}\ \emph {et~al.}(2005)\citenamefont {Bowman},
  \citenamefont {Heller}, \citenamefont {Leinweber}, \citenamefont
  {Parappilly}, \citenamefont {Williams},\ and\ \citenamefont
  {Zhang}}]{Bowman:2005vx}%
  \BibitemOpen
  \bibfield  {author} {\bibinfo {author} {\bibfnamefont {Patrick~O.}\
  \bibnamefont {Bowman}}, \bibinfo {author} {\bibfnamefont {Urs~M.}\
  \bibnamefont {Heller}}, \bibinfo {author} {\bibfnamefont {Derek~B.}\
  \bibnamefont {Leinweber}}, \bibinfo {author} {\bibfnamefont {Maria~B.}\
  \bibnamefont {Parappilly}}, \bibinfo {author} {\bibfnamefont {Anthony~G.}\
  \bibnamefont {Williams}}, \ and\ \bibinfo {author} {\bibfnamefont {Jian-Bo}\
  \bibnamefont {Zhang}},\ }\bibfield  {title} {\enquote {\bibinfo {title}
  {{Unquenched quark propagator in Landau gauge}},}\ }\href {\doibase
  10.1103/PhysRevD.71.054507} {\bibfield  {journal} {\bibinfo  {journal} {Phys.
  Rev.}\ }\textbf {\bibinfo {volume} {D71}},\ \bibinfo {pages} {054507}
  (\bibinfo {year} {2005})},\ \Eprint {http://arxiv.org/abs/hep-lat/0501019}
  {arXiv:hep-lat/0501019 [hep-lat]} \BibitemShut {NoStop}%
\bibitem [{\citenamefont {Furui}\ and\ \citenamefont
  {Nakajima}(2005)}]{Furui:2005mp}%
  \BibitemOpen
  \bibfield  {author} {\bibinfo {author} {\bibfnamefont {Sadataka}\
  \bibnamefont {Furui}}\ and\ \bibinfo {author} {\bibfnamefont {Hideo}\
  \bibnamefont {Nakajima}},\ }\bibfield  {title} {\enquote {\bibinfo {title}
  {{Unqueched Kogut-Susskind quark propagator in lattice Landau gauge QCD}},}\
  }\href@noop {} {\  (\bibinfo {year} {2005})},\ \Eprint
  {http://arxiv.org/abs/hep-lat/0511045} {arXiv:hep-lat/0511045 [hep-lat]}
  \BibitemShut {NoStop}%
\bibitem [{\citenamefont {Bonnet}\ \emph {et~al.}(2002)\citenamefont {Bonnet},
  \citenamefont {Bowman}, \citenamefont {Leinweber}, \citenamefont {Williams},\
  and\ \citenamefont {Zhang}}]{Bonnet:2002ih}%
  \BibitemOpen
  \bibfield  {author} {\bibinfo {author} {\bibfnamefont {Frederic D.~R.}\
  \bibnamefont {Bonnet}}, \bibinfo {author} {\bibfnamefont {Patrick~O.}\
  \bibnamefont {Bowman}}, \bibinfo {author} {\bibfnamefont {Derek~B.}\
  \bibnamefont {Leinweber}}, \bibinfo {author} {\bibfnamefont {Anthony~G.}\
  \bibnamefont {Williams}}, \ and\ \bibinfo {author} {\bibfnamefont {Jian-Bo}\
  \bibnamefont {Zhang}} (\bibinfo {collaboration} {CSSM Lattice}),\ }\bibfield
  {title} {\enquote {\bibinfo {title} {{Overlap quark propagator in Landau
  gauge}},}\ }\href {\doibase 10.1103/PhysRevD.65.114503} {\bibfield  {journal}
  {\bibinfo  {journal} {Phys. Rev.}\ }\textbf {\bibinfo {volume} {D65}},\
  \bibinfo {pages} {114503} (\bibinfo {year} {2002})},\ \Eprint
  {http://arxiv.org/abs/hep-lat/0202003} {arXiv:hep-lat/0202003 [hep-lat]}
  \BibitemShut {NoStop}%
\bibitem [{\citenamefont {Zhang}\ \emph {et~al.}(2004)\citenamefont {Zhang},
  \citenamefont {Bowman}, \citenamefont {Leinweber}, \citenamefont {Williams},\
  and\ \citenamefont {Bonnet}}]{Zhang:2003faa}%
  \BibitemOpen
  \bibfield  {author} {\bibinfo {author} {\bibfnamefont {J.~B.}\ \bibnamefont
  {Zhang}}, \bibinfo {author} {\bibfnamefont {Patrick~O.}\ \bibnamefont
  {Bowman}}, \bibinfo {author} {\bibfnamefont {Derek~B.}\ \bibnamefont
  {Leinweber}}, \bibinfo {author} {\bibfnamefont {Anthony~G.}\ \bibnamefont
  {Williams}}, \ and\ \bibinfo {author} {\bibfnamefont {Frederic D.~R.}\
  \bibnamefont {Bonnet}} (\bibinfo {collaboration} {CSSM Lattice}),\ }\bibfield
   {title} {\enquote {\bibinfo {title} {{Scaling behavior of the overlap quark
  propagator in Landau gauge}},}\ }\href {\doibase 10.1103/PhysRevD.70.034505}
  {\bibfield  {journal} {\bibinfo  {journal} {Phys. Rev.}\ }\textbf {\bibinfo
  {volume} {D70}},\ \bibinfo {pages} {034505} (\bibinfo {year} {2004})},\
  \Eprint {http://arxiv.org/abs/hep-lat/0301018} {arXiv:hep-lat/0301018
  [hep-lat]} \BibitemShut {NoStop}%
\bibitem [{\citenamefont {Zhang}\ \emph {et~al.}(2005)\citenamefont {Zhang},
  \citenamefont {Bowman}, \citenamefont {Coad}, \citenamefont {Heller},
  \citenamefont {Leinweber},\ and\ \citenamefont {Williams}}]{Zhang:2004gv}%
  \BibitemOpen
  \bibfield  {author} {\bibinfo {author} {\bibfnamefont {J.~B.}\ \bibnamefont
  {Zhang}}, \bibinfo {author} {\bibfnamefont {Patrick~O.}\ \bibnamefont
  {Bowman}}, \bibinfo {author} {\bibfnamefont {Ryan~J.}\ \bibnamefont {Coad}},
  \bibinfo {author} {\bibfnamefont {Urs~M.}\ \bibnamefont {Heller}}, \bibinfo
  {author} {\bibfnamefont {Derek~B.}\ \bibnamefont {Leinweber}}, \ and\
  \bibinfo {author} {\bibfnamefont {Anthony~G.}\ \bibnamefont {Williams}},\
  }\bibfield  {title} {\enquote {\bibinfo {title} {{Quark propagator in Landau
  and Laplacian gauges with overlap fermions}},}\ }\href {\doibase
  10.1103/PhysRevD.71.014501} {\bibfield  {journal} {\bibinfo  {journal} {Phys.
  Rev.}\ }\textbf {\bibinfo {volume} {D71}},\ \bibinfo {pages} {014501}
  (\bibinfo {year} {2005})},\ \Eprint {http://arxiv.org/abs/hep-lat/0410045}
  {arXiv:hep-lat/0410045 [hep-lat]} \BibitemShut {NoStop}%
\bibitem [{\citenamefont {Kamleh}\ \emph {et~al.}(2005)\citenamefont {Kamleh},
  \citenamefont {Bowman}, \citenamefont {Leinweber}, \citenamefont {Williams},\
  and\ \citenamefont {Zhang}}]{Kamleh:2004aw}%
  \BibitemOpen
  \bibfield  {author} {\bibinfo {author} {\bibfnamefont {Waseem}\ \bibnamefont
  {Kamleh}}, \bibinfo {author} {\bibfnamefont {Patrick~O.}\ \bibnamefont
  {Bowman}}, \bibinfo {author} {\bibfnamefont {Derek~B.}\ \bibnamefont
  {Leinweber}}, \bibinfo {author} {\bibfnamefont {Anthony~G.}\ \bibnamefont
  {Williams}}, \ and\ \bibinfo {author} {\bibfnamefont {Jianbo}\ \bibnamefont
  {Zhang}},\ }\bibfield  {title} {\enquote {\bibinfo {title} {{The fat link
  irrelevant clover overlap quark propagator}},}\ }\href {\doibase
  10.1103/PhysRevD.71.094507} {\bibfield  {journal} {\bibinfo  {journal} {Phys.
  Rev.}\ }\textbf {\bibinfo {volume} {D71}},\ \bibinfo {pages} {094507}
  (\bibinfo {year} {2005})},\ \Eprint {http://arxiv.org/abs/hep-lat/0412022}
  {arXiv:hep-lat/0412022 [hep-lat]} \BibitemShut {NoStop}%
\bibitem [{\citenamefont {Kamleh}\ \emph {et~al.}(2007)\citenamefont {Kamleh},
  \citenamefont {Bowman}, \citenamefont {Leinweber}, \citenamefont {Williams},\
  and\ \citenamefont {Zhang}}]{Kamleh:2007ud}%
  \BibitemOpen
  \bibfield  {author} {\bibinfo {author} {\bibfnamefont {Waseem}\ \bibnamefont
  {Kamleh}}, \bibinfo {author} {\bibfnamefont {Patrick~O.}\ \bibnamefont
  {Bowman}}, \bibinfo {author} {\bibfnamefont {Derek~B.}\ \bibnamefont
  {Leinweber}}, \bibinfo {author} {\bibfnamefont {Anthony~G.}\ \bibnamefont
  {Williams}}, \ and\ \bibinfo {author} {\bibfnamefont {Jianbo}\ \bibnamefont
  {Zhang}},\ }\bibfield  {title} {\enquote {\bibinfo {title} {{Unquenching
  effects in the quark and gluon propagator}},}\ }\href {\doibase
  10.1103/PhysRevD.76.094501} {\bibfield  {journal} {\bibinfo  {journal} {Phys.
  Rev.}\ }\textbf {\bibinfo {volume} {D76}},\ \bibinfo {pages} {094501}
  (\bibinfo {year} {2007})},\ \Eprint {http://arxiv.org/abs/0705.4129}
  {arXiv:0705.4129 [hep-lat]} \BibitemShut {NoStop}%
\bibitem [{\citenamefont {Schr{\"o}ck}(2012)}]{Schrock:2011hq}%
  \BibitemOpen
  \bibfield  {author} {\bibinfo {author} {\bibfnamefont {Mario}\ \bibnamefont
  {Schr{\"o}ck}},\ }\bibfield  {title} {\enquote {\bibinfo {title} {{The
  chirally improved quark propagator and restoration of chiral symmetry}},}\
  }\href {\doibase 10.1016/j.physletb.2012.04.008} {\bibfield  {journal}
  {\bibinfo  {journal} {Phys. Lett.}\ }\textbf {\bibinfo {volume} {B711}},\
  \bibinfo {pages} {217--224} (\bibinfo {year} {2012})},\ \Eprint
  {http://arxiv.org/abs/1112.5107} {arXiv:1112.5107 [hep-lat]} \BibitemShut
  {NoStop}%
\bibitem [{\citenamefont {Blossier}\ \emph {et~al.}(2011)\citenamefont
  {Blossier}, \citenamefont {Boucaud}, \citenamefont {Brinet}, \citenamefont
  {De~Soto}, \citenamefont {Liu}, \citenamefont {Morenas}, \citenamefont
  {P\`ene}, \citenamefont {Petrov},\ and\ \citenamefont
  {Rodr\'{\i}guez-Quintero}}]{Blossier:2010vt}%
  \BibitemOpen
  \bibfield  {author} {\bibinfo {author} {\bibfnamefont {B.}~\bibnamefont
  {Blossier}}, \bibinfo {author} {\bibfnamefont {Ph.}\ \bibnamefont {Boucaud}},
  \bibinfo {author} {\bibfnamefont {M.}~\bibnamefont {Brinet}}, \bibinfo
  {author} {\bibfnamefont {F.}~\bibnamefont {De~Soto}}, \bibinfo {author}
  {\bibfnamefont {Z.}~\bibnamefont {Liu}}, \bibinfo {author} {\bibfnamefont
  {V.}~\bibnamefont {Morenas}}, \bibinfo {author} {\bibfnamefont
  {O.}~\bibnamefont {P\`ene}}, \bibinfo {author} {\bibfnamefont
  {K.}~\bibnamefont {Petrov}}, \ and\ \bibinfo {author} {\bibfnamefont
  {J.}~\bibnamefont {Rodr\'{\i}guez-Quintero}},\ }\bibfield  {title} {\enquote
  {\bibinfo {title} {{Renormalisation of quark propagators from twisted-mass
  lattice QCD at $N_f$=2}},}\ }\href {\doibase 10.1103/PhysRevD.83.074506}
  {\bibfield  {journal} {\bibinfo  {journal} {Phys. Rev.}\ }\textbf {\bibinfo
  {volume} {D83}},\ \bibinfo {pages} {074506} (\bibinfo {year} {2011})},\
  \Eprint {http://arxiv.org/abs/1011.2414} {arXiv:1011.2414 [hep-ph]}
  \BibitemShut {NoStop}%
\bibitem [{\citenamefont {Burger}\ \emph {et~al.}(2013)\citenamefont {Burger}
  \emph {et~al.}}]{Burger:2012ti}%
  \BibitemOpen
  \bibfield  {author} {\bibinfo {author} {\bibfnamefont {Florian}\ \bibnamefont
  {Burger}} \emph {et~al.},\ }\bibfield  {title} {\enquote {\bibinfo {title}
  {{Quark mass and chiral condensate from the Wilson twisted mass lattice quark
  propagator}},}\ }\href {\doibase 10.1103/PhysRevD.87.034514,
  10.1103/PhysRevD.87.079904} {\bibfield  {journal} {\bibinfo  {journal} {Phys.
  Rev.}\ }\textbf {\bibinfo {volume} {D87}},\ \bibinfo {pages} {034514}
  (\bibinfo {year} {2013})},\ \bibinfo {note} {[Phys. Rev.D87,079904(2013)]},\
  \Eprint {http://arxiv.org/abs/1210.0838} {arXiv:1210.0838 [hep-lat]}
  \BibitemShut {NoStop}%
\bibitem [{\citenamefont {August}\ and\ \citenamefont
  {Maas}(2013)}]{August:2013jia}%
  \BibitemOpen
  \bibfield  {author} {\bibinfo {author} {\bibfnamefont {Daniel}\ \bibnamefont
  {August}}\ and\ \bibinfo {author} {\bibfnamefont {Axel}\ \bibnamefont
  {Maas}},\ }\bibfield  {title} {\enquote {\bibinfo {title} {{On the
  Landau-gauge adjoint quark propagator}},}\ }\href {\doibase
  10.1007/JHEP07(2013)001} {\bibfield  {journal} {\bibinfo  {journal} {JHEP}\
  }\textbf {\bibinfo {volume} {07}},\ \bibinfo {pages} {001} (\bibinfo {year}
  {2013})},\ \Eprint {http://arxiv.org/abs/1304.4423} {arXiv:1304.4423
  [hep-lat]} \BibitemShut {NoStop}%
\bibitem [{\citenamefont {Oliveira}\ \emph {et~al.}(2016)\citenamefont
  {Oliveira}, \citenamefont {K{\i}z{\i}lers{\"u}}, \citenamefont {Silva},
  \citenamefont {Skullerud}, \citenamefont {Sternbeck},\ and\ \citenamefont
  {Williams}}]{Oliveira:2016muq}%
  \BibitemOpen
  \bibfield  {author} {\bibinfo {author} {\bibfnamefont {Orlando}\ \bibnamefont
  {Oliveira}}, \bibinfo {author} {\bibfnamefont {Ay{\c s}e}\ \bibnamefont
  {K{\i}z{\i}lers{\"u}}}, \bibinfo {author} {\bibfnamefont {Paulo~J.}\
  \bibnamefont {Silva}}, \bibinfo {author} {\bibfnamefont {Jon-Ivar}\
  \bibnamefont {Skullerud}}, \bibinfo {author} {\bibfnamefont {Andre}\
  \bibnamefont {Sternbeck}}, \ and\ \bibinfo {author} {\bibfnamefont
  {Anthony~G.}\ \bibnamefont {Williams}},\ }\bibfield  {title} {\enquote
  {\bibinfo {title} {{Lattice Landau gauge quark propagator and the quark-gluon
  vertex}},}\ }\bibfield  {booktitle} {\emph {\bibinfo {booktitle}
  {{Proceedings, International Meeting Excited QCD 2016}}},\ }\href {\doibase
  10.5506/APhysPolBSupp.9.363} {\bibfield  {journal} {\bibinfo  {journal} {Acta
  Phys. Polon. Supp.}\ }\textbf {\bibinfo {volume} {9}},\ \bibinfo {pages}
  {363--368} (\bibinfo {year} {2016})},\ \Eprint
  {http://arxiv.org/abs/1605.09632} {arXiv:1605.09632 [hep-lat]} \BibitemShut
  {NoStop}%
\bibitem [{\citenamefont {Bali}\ \emph {et~al.}(2013)\citenamefont {Bali} \emph
  {et~al.}}]{Bali:2012qs}%
  \BibitemOpen
  \bibfield  {author} {\bibinfo {author} {\bibfnamefont {G.~S.}\ \bibnamefont
  {Bali}} \emph {et~al.},\ }\bibfield  {title} {\enquote {\bibinfo {title}
  {{Nucleon mass and sigma term from lattice QCD with two light fermion
  flavors}},}\ }\href {\doibase 10.1016/j.nuclphysb.2012.08.009} {\bibfield
  {journal} {\bibinfo  {journal} {Nucl. Phys.}\ }\textbf {\bibinfo {volume}
  {B866}},\ \bibinfo {pages} {1--25} (\bibinfo {year} {2013})},\ \Eprint
  {http://arxiv.org/abs/1206.7034} {arXiv:1206.7034 [hep-lat]} \BibitemShut
  {NoStop}%
\bibitem [{\citenamefont {Bali}\ \emph {et~al.}(2014)\citenamefont {Bali} \emph
  {et~al.}}]{Bali:2014gha}%
  \BibitemOpen
  \bibfield  {author} {\bibinfo {author} {\bibfnamefont {Gunnar~S.}\
  \bibnamefont {Bali}} \emph {et~al.},\ }\bibfield  {title} {\enquote {\bibinfo
  {title} {{The moment $\langle x\rangle_{u-d}$ of the nucleon from $N_f=2$
  lattice QCD down to nearly physical quark masses}},}\ }\href {\doibase
  10.1103/PhysRevD.90.074510} {\bibfield  {journal} {\bibinfo  {journal} {Phys.
  Rev.}\ }\textbf {\bibinfo {volume} {D90}},\ \bibinfo {pages} {074510}
  (\bibinfo {year} {2014})},\ \Eprint {http://arxiv.org/abs/1408.6850}
  {arXiv:1408.6850 [hep-lat]} \BibitemShut {NoStop}%
\bibitem [{\citenamefont {Bali}\ \emph {et~al.}(2015)\citenamefont {Bali} \emph
  {et~al.}}]{Bali:2014nma}%
  \BibitemOpen
  \bibfield  {author} {\bibinfo {author} {\bibfnamefont {Gunnar~S.}\
  \bibnamefont {Bali}} \emph {et~al.},\ }\href {\doibase
  10.1103/PhysRevD.91.054501} {\bibfield  {journal} {\bibinfo  {journal} {Phys.
  Rev.}\ }\textbf {\bibinfo {volume} {D91}},\ \bibinfo {pages} {054501}
  (\bibinfo {year} {2015})},\ \Eprint {http://arxiv.org/abs/1412.7336}
  {arXiv:1412.7336 [hep-lat]} \BibitemShut {NoStop}%
\bibitem [{\citenamefont {Sheikholeslami}\ and\ \citenamefont
  {Wohlert}(1985)}]{Sheikholeslami:1985ij}%
  \BibitemOpen
  \bibfield  {author} {\bibinfo {author} {\bibfnamefont {B.}~\bibnamefont
  {Sheikholeslami}}\ and\ \bibinfo {author} {\bibfnamefont {R.}~\bibnamefont
  {Wohlert}},\ }\bibfield  {title} {\enquote {\bibinfo {title} {{Improved
  Continuum Limit Lattice Action for QCD with Wilson Fermions}},}\ }\href
  {\doibase 10.1016/0550-3213(85)90002-1} {\bibfield  {journal} {\bibinfo
  {journal} {Nucl. Phys.}\ }\textbf {\bibinfo {volume} {B259}},\ \bibinfo
  {pages} {572} (\bibinfo {year} {1985})}\BibitemShut {NoStop}%
\bibitem [{\citenamefont {Heatlie}\ \emph {et~al.}(1991)\citenamefont
  {Heatlie}, \citenamefont {Martinelli}, \citenamefont {Pittori}, \citenamefont
  {Rossi},\ and\ \citenamefont {Sachrajda}}]{Heatlie:1990kg}%
  \BibitemOpen
  \bibfield  {author} {\bibinfo {author} {\bibfnamefont {G.}~\bibnamefont
  {Heatlie}}, \bibinfo {author} {\bibfnamefont {G.}~\bibnamefont {Martinelli}},
  \bibinfo {author} {\bibfnamefont {C.}~\bibnamefont {Pittori}}, \bibinfo
  {author} {\bibfnamefont {G.~C.}\ \bibnamefont {Rossi}}, \ and\ \bibinfo
  {author} {\bibfnamefont {Christopher~T.}\ \bibnamefont {Sachrajda}},\
  }\bibfield  {title} {\enquote {\bibinfo {title} {{The improvement of hadronic
  matrix elements in lattice QCD}},}\ }\href {\doibase
  10.1016/0550-3213(91)90137-M} {\bibfield  {journal} {\bibinfo  {journal}
  {Nucl. Phys.}\ }\textbf {\bibinfo {volume} {B352}},\ \bibinfo {pages}
  {266--288} (\bibinfo {year} {1991})}\BibitemShut {NoStop}%
\bibitem [{\citenamefont {Pennington}(2005)}]{Pennington:2005be}%
  \BibitemOpen
  \bibfield  {author} {\bibinfo {author} {\bibfnamefont {M.~R.}\ \bibnamefont
  {Pennington}},\ }\bibfield  {title} {\enquote {\bibinfo {title} {{Swimming
  with quarks}},}\ }\bibfield  {booktitle} {\emph {\bibinfo {booktitle}
  {{Particles and fields. Proceedings, 11th Mexican School, Xalapa, Veracruz,
  Mexico, August 2-13, 2004}}},\ }\href {\doibase 10.1088/1742-6596/18/1/001}
  {\bibfield  {journal} {\bibinfo  {journal} {J. Phys. Conf. Ser.}\ }\textbf
  {\bibinfo {volume} {18}},\ \bibinfo {pages} {1--73} (\bibinfo {year}
  {2005})},\ \Eprint {http://arxiv.org/abs/hep-ph/0504262}
  {arXiv:hep-ph/0504262 [hep-ph]} \BibitemShut {NoStop}%
\bibitem [{\citenamefont {Leinweber}\ \emph {et~al.}(1999)\citenamefont
  {Leinweber}, \citenamefont {Skullerud}, \citenamefont {Williams},\ and\
  \citenamefont {Parrinello}}]{Leinweber:1998uu}%
  \BibitemOpen
  \bibfield  {author} {\bibinfo {author} {\bibfnamefont {Derek~B.}\
  \bibnamefont {Leinweber}}, \bibinfo {author} {\bibfnamefont {Jon~Ivar}\
  \bibnamefont {Skullerud}}, \bibinfo {author} {\bibfnamefont {Anthony~G.}\
  \bibnamefont {Williams}}, \ and\ \bibinfo {author} {\bibfnamefont {Claudio}\
  \bibnamefont {Parrinello}} (\bibinfo {collaboration} {UKQCD}),\ }\bibfield
  {title} {\enquote {\bibinfo {title} {{Asymptotic scaling and infrared
  behavior of the gluon propagator}},}\ }\href {\doibase
  10.1103/PhysRevD.61.079901, 10.1103/PhysRevD.60.094507} {\bibfield  {journal}
  {\bibinfo  {journal} {Phys. Rev.}\ }\textbf {\bibinfo {volume} {D60}},\
  \bibinfo {pages} {094507} (\bibinfo {year} {1999})},\ \bibinfo {note}
  {[Erratum: Phys. Rev.D61,079901(2000)]},\ \Eprint
  {http://arxiv.org/abs/hep-lat/9811027} {arXiv:hep-lat/9811027 [hep-lat]}
  \BibitemShut {NoStop}%
\bibitem [{\citenamefont {Becirevic}\ \emph {et~al.}(1999)\citenamefont
  {Becirevic}, \citenamefont {Boucaud}, \citenamefont {Leroy}, \citenamefont
  {Micheli}, \citenamefont {P{\`e}ne}, \citenamefont
  {Rodr{\'\i}guez-Quintero},\ and\ \citenamefont
  {Roiesnel}}]{Becirevic:1999uc}%
  \BibitemOpen
  \bibfield  {author} {\bibinfo {author} {\bibfnamefont {D.}~\bibnamefont
  {Becirevic}}, \bibinfo {author} {\bibfnamefont {Philippe}\ \bibnamefont
  {Boucaud}}, \bibinfo {author} {\bibfnamefont {J.~P.}\ \bibnamefont {Leroy}},
  \bibinfo {author} {\bibfnamefont {J.}~\bibnamefont {Micheli}}, \bibinfo
  {author} {\bibfnamefont {O.}~\bibnamefont {P{\`e}ne}}, \bibinfo {author}
  {\bibfnamefont {J.}~\bibnamefont {Rodr{\'\i}guez-Quintero}}, \ and\ \bibinfo
  {author} {\bibfnamefont {C.}~\bibnamefont {Roiesnel}},\ }\bibfield  {title}
  {\enquote {\bibinfo {title} {{Asymptotic behavior of the gluon propagator
  from lattice QCD}},}\ }\href {\doibase 10.1103/PhysRevD.60.094509} {\bibfield
   {journal} {\bibinfo  {journal} {Phys. Rev.}\ }\textbf {\bibinfo {volume}
  {D60}},\ \bibinfo {pages} {094509} (\bibinfo {year} {1999})},\ \Eprint
  {http://arxiv.org/abs/hep-ph/9903364} {arXiv:hep-ph/9903364 [hep-ph]}
  \BibitemShut {NoStop}%
\bibitem [{\citenamefont {de~Soto}\ and\ \citenamefont
  {Roiesnel}(2007)}]{deSoto:2007ht}%
  \BibitemOpen
  \bibfield  {author} {\bibinfo {author} {\bibfnamefont {F.}~\bibnamefont
  {de~Soto}}\ and\ \bibinfo {author} {\bibfnamefont {C.}~\bibnamefont
  {Roiesnel}},\ }\bibfield  {title} {\enquote {\bibinfo {title} {{On the
  reduction of hypercubic lattice artifacts}},}\ }\href {\doibase
  10.1088/1126-6708/2007/09/007} {\bibfield  {journal} {\bibinfo  {journal}
  {JHEP}\ }\textbf {\bibinfo {volume} {09}},\ \bibinfo {pages} {007} (\bibinfo
  {year} {2007})},\ \Eprint {http://arxiv.org/abs/0705.3523} {arXiv:0705.3523
  [hep-lat]} \BibitemShut {NoStop}%
\bibitem [{\citenamefont {Fischer}\ and\ \citenamefont
  {Alkofer}(2003)}]{Fischer:2003rp}%
  \BibitemOpen
  \bibfield  {author} {\bibinfo {author} {\bibfnamefont {Christian~S.}\
  \bibnamefont {Fischer}}\ and\ \bibinfo {author} {\bibfnamefont {Reinhard}\
  \bibnamefont {Alkofer}},\ }\bibfield  {title} {\enquote {\bibinfo {title}
  {{Nonperturbative propagators, running coupling and dynamical quark mass of
  Landau gauge QCD}},}\ }\href {\doibase 10.1103/PhysRevD.67.094020} {\bibfield
   {journal} {\bibinfo  {journal} {Phys. Rev.}\ }\textbf {\bibinfo {volume}
  {D67}},\ \bibinfo {pages} {094020} (\bibinfo {year} {2003})},\ \Eprint
  {http://arxiv.org/abs/hep-ph/0301094} {arXiv:hep-ph/0301094 [hep-ph]}
  \BibitemShut {NoStop}%
\bibitem [{\citenamefont {Aguilar}\ and\ \citenamefont
  {Papavassiliou}(2011)}]{Aguilar:2010cn}%
  \BibitemOpen
  \bibfield  {author} {\bibinfo {author} {\bibfnamefont {A.~C.}\ \bibnamefont
  {Aguilar}}\ and\ \bibinfo {author} {\bibfnamefont {J.}~\bibnamefont
  {Papavassiliou}},\ }\bibfield  {title} {\enquote {\bibinfo {title} {{Chiral
  symmetry breaking with lattice propagators}},}\ }\href {\doibase
  10.1103/PhysRevD.83.014013} {\bibfield  {journal} {\bibinfo  {journal} {Phys.
  Rev.}\ }\textbf {\bibinfo {volume} {D83}},\ \bibinfo {pages} {014013}
  (\bibinfo {year} {2011})},\ \Eprint {http://arxiv.org/abs/1010.5815}
  {arXiv:1010.5815 [hep-ph]} \BibitemShut {NoStop}%
\bibitem [{\citenamefont {Constantinou}\ \emph {et~al.}(2009)\citenamefont
  {Constantinou}, \citenamefont {Lubicz}, \citenamefont {Panagopoulos},\ and\
  \citenamefont {Stylianou}}]{Constantinou:2009tr}%
  \BibitemOpen
  \bibfield  {author} {\bibinfo {author} {\bibfnamefont {M.}~\bibnamefont
  {Constantinou}}, \bibinfo {author} {\bibfnamefont {V.}~\bibnamefont
  {Lubicz}}, \bibinfo {author} {\bibfnamefont {H.}~\bibnamefont
  {Panagopoulos}}, \ and\ \bibinfo {author} {\bibfnamefont {F.}~\bibnamefont
  {Stylianou}},\ }\bibfield  {title} {\enquote {\bibinfo {title} {{$O(a^2)$
  corrections to the one-loop propagator and bilinears of clover fermions with
  Symanzik improved gluons}},}\ }\href {\doibase 10.1088/1126-6708/2009/10/064}
  {\bibfield  {journal} {\bibinfo  {journal} {JHEP}\ }\textbf {\bibinfo
  {volume} {10}},\ \bibinfo {pages} {064} (\bibinfo {year} {2009})},\ \Eprint
  {http://arxiv.org/abs/0907.0381} {arXiv:0907.0381 [hep-lat]} \BibitemShut
  {NoStop}%
\bibitem [{\citenamefont {G{\"o}ckeler}\ \emph {et~al.}(2010)\citenamefont
  {G{\"o}ckeler} \emph {et~al.}}]{Gockeler:2010yr}%
  \BibitemOpen
  \bibfield  {author} {\bibinfo {author} {\bibfnamefont {M.}~\bibnamefont
  {G{\"o}ckeler}} \emph {et~al.},\ }\bibfield  {title} {\enquote {\bibinfo
  {title} {{Perturbative and Nonperturbative Renormalization in Lattice
  QCD}},}\ }\href {\doibase 10.1103/PhysRevD.82.114511,
  10.1103/PhysRevD.86.099903} {\bibfield  {journal} {\bibinfo  {journal} {Phys.
  Rev.}\ }\textbf {\bibinfo {volume} {D82}},\ \bibinfo {pages} {114511}
  (\bibinfo {year} {2010})},\ \bibinfo {note} {[Erratum: Phys.
  Rev.D86,099903(2012)]},\ \Eprint {http://arxiv.org/abs/1003.5756}
  {arXiv:1003.5756 [hep-lat]} \BibitemShut {NoStop}%
\end{thebibliography}%

\end{document}